\documentstyle [12pt,epsfig]{article}
\makeatletter
\@addtoreset{equation}{section}
\makeatother

\newcommand{\BS}{\bigskip}
\newcommand{\SECTION}[1]{\BS{\large\section{\bf #1}}}

\newcommand{\KS}{{\rm K}_S}
\newcommand{\KL}{{\rm K}_L}
\newcommand{\BH}{{\rm B}_H}
\newcommand{\BL}{{\rm B}_L}
\begin{document}
\begin{titlepage}
\begin{center}
\vspace*{2cm}
{\large \bf A Covariant Path Amplitude Description of Flavour Oscillations:
 The Gribov-Pontecorvo Phase for Neutrino Vacuum Propagation 
 is Right}
\vspace*{1.5cm}
\end{center}
\begin{center}
{\bf J.H.Field }
\end{center}
\begin{center}
{ 
D\'{e}partement de Physique Nucl\'{e}aire et Corpusculaire
 Universit\'{e} de Gen\`{e}ve . 24, quai Ernest-Ansermet
 CH-1211 Gen\`{e}ve 4.
}
\end{center}
\vspace*{2cm}
\begin{abstract}
  An extended study is performed of geometrical and kinematical assumptions 
  used in calculations of the neutrino oscillation phase.
  The almost universally employed `equal velocity' assumption, in which all 
  neutrino mass eigenstates are produced at the same time, is shown to
  underestimate, by a factor of two, the neutrino propagation contribution
  to the phase. Taking properly into account, in a covariant path amplitude
  calculation, the incoherent nature of neutrino production as predicted
  by the Standard Model, results in an important source propagator contribution
   to the phase. It is argued that the commonly discussed Gaussian
  `wave packets' have no basis within quantum mechanics and are the result
   of a confused amalgam of quantum and classical wave concepts.                     
\end{abstract}
\vspace*{1cm}
\newline
{\it Keywords ;} Quantum Mechanics,
Neutrino Oscillations.
\end{titlepage}
  
\SECTION{\bf{Introduction}}
  The first published calculation of the phase of neutrino oscillations,
   on the assumption that neutrinos are massive, but ultra-relativistic, 
   was made in 1969 by Gribov and Pontecorvo~\cite{GribPont}. No details
   of the method used to obtain the phase were given. Later publications
   used this result for phenomenology. In 1976 a paper was published~\cite{FritMink}
   in which an oscillation phase a factor of two smaller than the
   Gribov-Pontecorvo result was obtained. No comment was made, then or later, on
   this discrepancy, which remained unnoticed before a paper written
   recently by the present author~\cite{JHF1} in which more bibliographic details
   of this `missing controversy' can be found. Since the first derivation
   of the `standard' oscillation phase, as the result of Reference~\cite{FritMink}
   will be referred to in this paper, of the order of a hundred papers have been
   published devoted to the quantum mechanics (QM) of neutrino oscillations.
   \par The present paper has three aims. The first is to carefully review the
   kinematical and geometrical assumptions of previous calculations, particularly
   with regard to their consistency in terms of an expansion in powers of
   the neutrino masses. The second is to present a calculation of the oscillation
   phase, for sources at rest, using a covariant Feynman path amplitude
   formulation~\cite{Feyn2,Feyn3} of QM
    \footnote{Feyman's original
   work on this subject cited in References~\cite{Feyn2,Feyn3} concerned only 
   non-relativistic QM, but the fundamental formula (5.1), on which the work
   presented in the present paper is based, holds also in relativistic QM. The
    approach should be more properly termed a `path amplitude' rather than
   a `path integral' one as no attempt is made to evaluate the 
   latter in closed mathemetical form, as is done, for example, in the
    discussion of bound state problems.}   
  The third is to critically discuss
   treatments, in the literature, of the QM of neutrino oscillations, in the
   light of both the kinematical and geometrical findings of the present paper
   and the physical insights provided by the path amplitude treatment.
   \par The calculations presented here are only a sub-set of those to be found in
    a previous paper by the present author~\cite{JHF1}, results of which are also
    summarised in a short letter~\cite{JHF2}. The calculations in
    Reference~\cite{JHF1} include neutrino oscillations following decay at rest
    pions, muons and $\beta$-unstable nuclei, muon oscillations following pion decay 
    at rest, and neutrino and muon oscillations following decay in flight of
    ultra-relativistic pions. Also discussed in Reference~\cite{JHF1}  are
    different mechanisms
    (both coherent and incoherent) that may contribute to damping of the
     oscillations. An attempt has been made, in the present paper, to 
     give a more pedagogical presentation in which the crucial underlying
    physical assumptions used are spelt out as clearly as possible.
    \par Reference~\cite{JHF1} contains already a extensive critical review of the
    previous literature. In the present paper, the further discussion has
    been largely motivated and orientated by the reports of several anonymous reviewers of
    References~\cite{JHF1,JHF2}. Since writing these papers I became aware of the
    important work of Shrock~\cite{Shrock1,Shrock2} who pointed out, more than 
    twenty years ago, the incoherent\footnote{The word `incoherent' here means
    that the different mass eigenstates are produced in independent physical
    processes, not in a coherent `flavour eigenstate' that is a superposition
    of the mass eigenstates. A `coherent source' is defined as one in which
    the amplitudes corresponding to different decay times have a well-defined
    phase difference.}
    nature of the weak decay processes in
    which neutrino mass eigenstates are produced. This was indeed one of
    the crucial assumptions for the calculations presented in
    References~\cite{JHF1,JHF2}. I also now fully understand the (simple)
    reason why the standard oscillation phase has been obtained in essentially
    all published derivations since Reference~\cite{FritMink}. Indeed, as will be
    demonstrated, the wrong (standard) instead of the correct (Gribov-Pontecorvo)
    neutrino propagation phase has been universally obtained due to the universal
    failure to use the correct description of the weak neutrino production
    processes as formulated by Shrock~\cite{Shrock1,Shrock2}. 
    \par For simplicity, only two-flavour mixing will be considered, so that the
    mixing amplitudes: $\langle i | \alpha \rangle$ (where $ i~(=~1,2)$ refers to 
    neutrino mass eigenstates $\nu_1,\nu_2$ and $\alpha~(=~e,\mu)$ refers to the
    flavour of the charged lepton participating in the weak interaction process)
    may be taken to be real. Unitarity then enables all mixing
    amplitudes to be described in terms of a single angular parameter, $\theta$,
    as shown in Table 1.   
   \begin{table}
   \begin{center}
   \begin{tabular}{|c|c c c c|} \hline  
       $p$  & 1  & 2  & e & $\mu$  \\
   \hline \hline
      $q$  &   &   &  &  \\ \cline{1-1}
     1  & 0  & 0  & $ \cos \theta$  & $-\sin \theta$  \\
     2 & 0  & 0  & $ \sin \theta$  & $\cos\theta$  \\
     e & $ \cos \theta$  & $\sin \theta$ & 0 & 0   \\
     $\mu$ & $ -\sin \theta$  & $\cos \theta$ & 0 & 0   \\ 
  \hline
  \end{tabular}
   \caption[]{{\it Values of the flavour-mass mixing 
    amplitude $\langle p | q \rangle$ }}
  \end{center}
  \end{table}
  Note that there are no `neutrino flavour eigenstates'. The amplitude
  $\langle 1 | \mu \rangle$, for example, is the amplitude to produce
  the neutrino mass eigenstate $\nu_1$ in association with (or in the decay of)
  a muon. Thus the flavour-mass mixing amplitudes fix the strength of
   the charged current couplings to the  neutrino mass eigenstates~\cite{Shrock1,Shrock2}.
    For example, in the decays: $\pi^+,~\rm{K}^+,~\rm{B}^+ \rightarrow \mu^+ \nu_i$
   the leptonic charged current is:
   $\overline{\nu_i}\gamma_{\rho}(1-\gamma_5)U_{\mu i}\mu$ ($i=1,2$) to be compared
   with the quark currents:  $\overline{q}\gamma_{\rho}(1-\gamma_5)V_{uq}u$
   ($q=d,s,b$),
   where $U_{\mu i} = U_{i \mu} \equiv \langle 1 | \mu \rangle$ is an element of the
   Maki-Nakagawa-Sakata (MNS)~\cite{MNS} lepton flavour/mass mixing matrix, and 
   $V_{uq}$ is an element  of the Cabibbo-Kobayashi-Maskawa (CKM)~\cite{CKM}
   flavour/mass mixing matrix in the quark sector.   
   The couplings of antineutrinos to charged anti-leptons are the same
   as those of neutrinos to charged leptons in the two flavour mixing case.
   \par The amplitude to produce a charged lepton of flavour $\beta$ by the interaction
   of a neutrino mass eigenstate $i$ produced in a decay involving
   a charged lepton\footnote{In the decay process, this charged lepton,
    together with the neutrino $\nu_i$, form a charged current that couples to
    a real or virtual W boson.} of flavour $\alpha$ may, 
   in general, be written as:
   \begin{equation}
    A_i(\beta \leftarrow \alpha) = A_0 \exp(-i\Delta \phi_i)\langle \beta | i \rangle
    \langle i | \alpha \rangle
   \end{equation}
    The probability to observe a charged lepton of flavour $\beta$, in the case that the
     neutrino mass eigenstates are not distinguished, is then, by superposition:
    \begin{eqnarray}
    P_{\beta \alpha} & = & |A_1(\beta \leftarrow \alpha) + A_2(\beta \leftarrow \alpha)|^2
    \nonumber \\
   & = & |A_0|^2\left[(\langle \beta |1 \rangle  \langle 1 |  \alpha \rangle)^2
         + (\langle \beta | 2 \rangle  \langle 2  |  \alpha \rangle)^2\right.
   \nonumber \\
   &  & \left. + 2 \langle \beta |1 \rangle \langle \beta | 2 \rangle \langle 1 | 
 \alpha \rangle \langle 2  |  \alpha \rangle \cos(\Delta \phi_1-\Delta \phi_2) \right]
   \end{eqnarray}
    The `neutrino oscillation phase' $\phi_{12}$ is defined as 
   \begin{equation}
    \phi_{12} \equiv \Delta \phi_1-\Delta \phi_2 
   \end{equation}
 The Gribov-Pomeranchuk value of the oscillation phase is:
  \begin{equation}
    \phi_{12}^{GP} = \frac{\Delta m^2 L}{p_{\nu}} 
   \end{equation}
 whereas the standard result is:
  \begin{equation}
    \phi_{12}^{stand} = \frac{\Delta m^2 L}{2 p_{\nu}} 
   \end{equation}
 In Eqns(1.4) and (1.5), $\Delta m^2 \equiv m_1^2-m_2^2$ where $m_1$ and $m_2$ are
  the neutrino masses, $L$ is the source-detector distance and and $p_{\nu}$ is the 
   measured neutrino momentum. Units with $\hbar = c =1$ are used throughout 
   this paper. The word `neutrino', without further qualification, stands for
   `neutrino mass eigenstate'.
 The plan of the paper is as follows: In Sections 2, 3 and 4 various
   kinematical and geometrical approximations that have been used in neutrino
   oscillation calculations are discussed. In particular it will be
   examined whether particular approximations retain, or not, all the leading
  order O($m^2$) terms, or contribute only negligible O($m^4$) corrections.
  The Lorentz invariant nature of the oscillation phase is fully exploited in
  these purely mathematical considerations. Section 5 describes the calculation
  of the oscillation phase using the covariant Feynman path amplitude method.
  Section 6 is devoted to a discussion of previous treatments of the QM
  of neutrino oscillations. In particular, the physical basis of widely used
   Gaussian wave packet models is questioned. Section 7 contains a brief 
   summary and outlook.
         
\SECTION{\bf{Lorentz Invariant Plane Wave Propagation}}
 In this case, the one-dimensional propagation amplitude for a particle
 of mass $m_i$ over space and time intervals: $\Delta x_i$ and $\Delta t_i$
 is given by:
\begin{eqnarray}
 P(\Delta x_i,\Delta t_i, m_i) & = & P_0 \exp\left\{-i[E_i\Delta t_i-p_i\Delta x_i]\right\}   
 \nonumber \\
 & = &  P_0 \exp\left\{-im_i \Delta \tau_i\right\} \nonumber \\
 &\equiv &  P_0 \exp\left\{-i\Delta \phi_i\right\}
\end{eqnarray} 
Here $\Delta \tau_i$ is the proper time interval, in the rest frame of the particle, $i$,
 corresponding to space and time intervals $\Delta x_i$ and $\Delta t_i$ in the laboratory
 frame, so that:
\begin{equation}
 \Delta \tau_i^2 = (\Delta t_i)^2 - (\Delta x_i)^2  
\end{equation}
 The relativistic relation between energy, momentum and mass:
\begin{equation}
 E_i(p_i,m_i) = \sqrt{p_i^2+m_i^2}
\end{equation}
 implies a group velocity $v^G_i$ for the travelling wave represented by Eqn(2.1)
 given by:
\begin{equation}
 v^G_i \equiv \frac {dE_i}{dp_i} = \frac{p_i}{E_i}
\end{equation}
 In the following, it is assumed that the neutrino mass eigenstates whose
  propagation is described by
 Eqn(2.1), have velocity $v_i = v^G_i$. As correctly emphasised in
  Reference~\cite{Lipkin1} particle oscillation experiments
  actually measure the spatial, not the temporal, dependence of
  quantum mechanical interference effects. Therefore, throughout
  the kinematical discussions in the present and following Sections,
  a fixed distance, $\Delta x_i = L$, is assumed between the source 
  particle (at rest) and the detection event. With the additional 
  assumption that the propagating neutrino mass eigenstates are on-shell 
  particles,
  several different exact expressions may be written for the
   Lorentz-invariant phase increment:
\begin{equation}
 \Delta \phi_i = m_i \Delta \tau_i = \frac{m_i \Delta t_i}{\gamma_i}
  = \frac{m_i^2 \Delta t_i}{E_i} = \frac{m_i^2 L}{p_i}
\end{equation}
 where, in the last member of Eqn(2.5), the relation: $\Delta t_i = L/v_i$
 has been used, and $\gamma = 1/\sqrt{1-v^2}= E/m$ is the usual parameter of
 Special Relativity.  
 \par The value of the oscillation phase: $\phi_{12} \equiv \Delta \phi_1
  - \Delta \phi_2$ obtained using different kinematical approximations will 
  now be considered. For this it is useful to introduce the quantity
  $p_0$ which is the limiting value of $p_i$ or $E_i$ as 
   $m_i \rightarrow 0$. The following
    expressions for $p_i$ and $E_i$ may then be written:
\begin{eqnarray}
 p_i & = & p_0\left\{1 - \left(\frac{m_i}{m_S}\right)^2\left[\frac{1+R_m^2}
 {(1-R_m^2)^2}\right]\right\}+ O(m^4)   \\ 
 E_i & = & p_0\left\{1 + \left(\frac{m_i}{m_S}\right)^2\left[\frac{1}
 {1-R_m^2}\right]\right\}
\end{eqnarray}
 where 
\begin{equation}
p_0 = \frac{m_S}{2}(1-R_m^2)
\end{equation}
Here  $m_S$ is the mass of the decaying source particle and
 $R_m \equiv m_R/m_S$ where $m_R$ is the mass of the particle
 (or system of particles) recoiling against the neutrino in the
 decay process. It is also convenient to introduce the average kinematical
 quantities for the neutrino mass eigenstates:
\begin{eqnarray}
 \overline{m} & = & \frac{(m_1+m_2)}{2}  \\
 \overline{p} & = & \frac{(p_1+p_2)}{2}  \\
 \overline{E} & = &  \sqrt{\overline{p}^2+\overline{m}^2}
\end{eqnarray}  
If, as is usually done in the literature, common values of both
 $\Delta x$ and $\Delta t$ are assumed for both mass eigenstates,
 then:
\begin{equation}
v_1 = v_2 = \overline{v} = \frac{\Delta x}{\Delta t} = \frac{\overline{p}}
{\overline{E}}
\end{equation} 
and   
\begin{equation}
\gamma_1 = \gamma_2 = \overline{\gamma} = \frac{\Delta t}{\Delta \tau} = 
\frac{ \overline{E}}{\overline{m}}
\end{equation}
 It must be remarked at once that such an `equal velocity' hypothesis
 is in contradiction with the fundamental definitions in Eqns(2.1)-(2.4).
 For logical consistency, if $\Delta x$ and $\Delta t$ are assumed to be 
  equal for the two mass eigenstates, then also $E_i$ and $p_i$ in Eqn(2.1)
 should be replaced by $\overline{E}$ and $\overline{p}$. Then, evidently,
 no neutrino oscillations would be possible. In fact the equal velocity 
 hypothesis assumes not only Eqn(2.12) but also that the energy and momentum
  appearing in the 4-vector product in (2.1) are in accordance with 
  energy momentum conservation in the decay process. Since
    $v_i^G = p_i/E_i \ne \overline{v}$, there is then an internal
   contradiction. Also,
   because of Eqn(2.12), the corresponding space and time intervals
   do not respect the equations: $v_i^G = \Delta x_i/\Delta t_i$, i.e.
   the neutrinos do not move along classical trajectories 
   specified by the group velocities $v_i^G$. Thus the equal velocity
   hypothesis respects energy-momentum conservation, but not the 
   constraints of space-time geometry, and contains an internal contradiction.
   The formula (2.1) combines the quantities ($\Delta x_i$, $\Delta t_i$)
 from space-time geometry and ($E_i$, $p_i$) from particle kinematics. The equal 
  velocity hypothesis then treats, in an inconsistent manner, these two sets of
  quantities.
  Introducing the notation $\Delta \phi_i(\overline{v})$ for the approximate phase
 increment calculated with fixed velocity $\overline{v}$, according to
 Eqns(2.9)-(2.13), then:
\begin{equation}
 \Delta \phi_i(\overline{v}) = m_i \Delta \tau = \frac{m_i \Delta t}
 {\overline{\gamma}} =
  \frac{m_i^2 \Delta t}{\overline{E}} = \frac{m_i^2 L}{\overline{p}}
\end{equation}
 Now denoting by the superscripts $I$, $II$, $III$ and $IV$ the values of
 $\phi_{12}(\overline{v})$ calculated using the successive members of Eqn(2.14)
 the following results are obtained:
\begin{eqnarray}
 \phi_{12}^{I}(\overline{v}) & = & (m_1-m_2) \Delta \tau = (m_1-m_2)
 \frac{\overline{m} L}{\overline{p}} = \frac{(m_1^2-m_2^2)L}{2\overline{p}} \\
 \phi_{12}^{II}(\overline{v}) & = & (m_1-m_2) \frac{\Delta t}
 {\overline{\gamma}} = (m_1-m_2)
 \frac{\overline{m} L}{\overline{p}} = \frac{(m_1^2-m_2^2)L}{2\overline{p}} \\
 \phi_{12}^{III}(\overline{v}) & = & (m_1^2-m_2^2) \frac{\Delta t}
 {\overline{E}} = \frac{(m_1^2-m_2^2)L}{\overline{p}} \\
 \phi_{12}^{IV}(\overline{v}) & = & \frac{(m_1^2-m_2^2)L}{\overline{p}}
\end{eqnarray} 
 The phases $\phi_{12}^{III}$ and $\phi_{12}^{IV}$ are found to be equal and 
 a factor of two larger than either $\phi_{12}^{I}$ or $\phi_{12}^{II}$.
 The different results obtained indicate that some inconsistent
 approximations must have been made in the passage from the exact
 expressions for the phase increment in (2.5) to the approximate ones 
 in (2.14). To see just where the inconsistency has occured, and which results,
 $I$ and $II$ or $III$ and $IV$, are correct, the calculation is now done using the 
 exact formulae of Eqn(2.5) and, for ease of comparison with the approximate
 expressions in Eqn(2.14) the neutrino masses are written as:
\begin{equation}
 m_i = \overline{m}+\delta_i
\end{equation}
 where 
\begin{equation}
 \delta_1 = -\delta_2  = \frac{m_1-m_2}{2}
\end{equation}       
 Thus $\delta_i$ are correction terms, relating the approximate relations
 in Eqns(2.14) to the exact ones in Eqn(2.5). Denoting by $\phi_{12}(exact)$ the
  interference phase calculated using Eqn(2.5) then:
\begin{equation}
\phi_{12}(exact)  =  m_1 \Delta \tau_1- m_2 \Delta \tau_2
 = m_1 \frac{\Delta t_1}{\gamma_1} - m_2 \frac{\Delta t_2}{\gamma_2}
 = \left(\frac{m_1^2}{p_1}-\frac{m_2^2}{p_2}\right)L 
 = \frac{(m_1^2-m_2^2)}{\overline{p}}L + O(m^4) 
\end{equation} 
 Thus the result obtained is in agreement, at O($m^2$), with the approximate 
 calculations in cases III and IV. Using now Eqns(2.19) and (2.20):
\begin{eqnarray}
\phi_{12}(exact) &  =  & \left[\frac{(\overline{m}+\delta_1)^2}{p_1}-\frac{
 (\overline{m}+\delta_2)^2}{p_2}\right]L
  = \left[\frac{(\overline{m}+\delta_1)^2-
 (\overline{m}+\delta_2)^2}{\overline{p}}\right]L
  + O(m^4) \nonumber  \\
    & = & \frac{(m_1^2-m_2^2)}{2\overline{p}}L +
 \frac{(m_1-m_2)\overline{m}}{\overline{p}}L+ O(m^4)       
\end{eqnarray} 
 Comparing the second term on the RHS this equation with Eqns(2.15) or
 (2.16) 
 it can be seen that the correction terms in Eqn(2.19) have the effect of
 doubling the approximate results $\phi_{12}^{I}(\overline{v})$ and 
 $\phi_{12}^{II}(\overline{v})$. Thus important O($m^2$) contributions
 are neglected in Eqns(2.15) and (2.16). The reason why the
 constant velocity approximation is so poor in Eqns(2.15) and (2.16)
 becomes evident on inspection of Eqn(2.13). Unlike in the case of 
 neutral kaons or b-mesons, the neutrino masses
 may be widely different, so that although, for highly relativistic 
 neutrinos, the velocities of the mass eigenstates may be very similar,
 this is not the case for the Lorentz-$\gamma$ factors in Eqn(2.13) that
 are inversely proportional to the neutrino masses. The important difference
 between the exact and approximate calculations occurs in the very first
 member of Eqn(2.15). Indeed, as is shown by Eqns(2.21) and (2.15):
\begin{equation}
\phi_{12}(exact)-\phi_{12}^{I}(\overline{v}) =  m_1 \Delta \tau_1- m_2 \Delta \tau_2
 - (m_1-m_2) \Delta \tau =  \frac{(m_1^2-m_2^2)}{2 \overline{p}}L + O(m^4)
\end{equation}  
 In fact, in Eqns(2.17) and (2.18), the approximate Lorentz-$\gamma$ factor
 $\overline{\gamma} = \overline{E}/\overline{m}$ is replaced by the
 still approximate but mass dependent $\gamma$-factors
  $\overline{E}/m_i$
 which differ from the
 exact ones $E_i/m_i$ by corrections of only O($m^2$). Thus, up to corrections
 of O($m^4$), the same result is obtained as the exact formula. On the other
 hand the difference between $1/\overline{\gamma}$, used in Eqn(2.16), and $1/\gamma_1$ or 
 $1/\gamma_2$ used in Eqn(2.5), is of O($m$) and so cannot be neglected in an O($m^2$)
 calculation of the oscillation phase. 
 \par It is instructive to calculate the oscillation phase directly in the
 laboratory system, using Eqn(2.1), rather than by Lorentz transformation
 from the neutrino center of mass to the laboratory system as in Eqns(2.5)
 and (2.14). Considering first the equal velocity case:
\begin{eqnarray}
\phi_{12}^{LAB}(\overline{v}) &  =  & (E_1-E_2)\Delta t-(p_1-p_2)L \nonumber \\
      & = & [(E_1-E_2)\frac{1}{\overline{v}}-(p_1-p_2)]L    
\end{eqnarray}
Using the relations:
\begin{eqnarray}
E_i &  =  & \sqrt{p_i^2+m_i^2} = p_i +\frac{m_i^2}{2p_i} + O(m^4) \\
\overline{v} & = & 1-\frac{\overline{m}^2}{2\overline{p}^2} + O(m^4)     
\end{eqnarray} 
 and noting that, from Eqns(2.6) and (2.10) that $p_i = \overline{p}+O(m^2)$, Eqn(2.24) may 
 be written as:
\begin{eqnarray}
\phi_{12}^{LAB}(\overline{v}) &  =  & \left(\frac{m_1^2}{2p_1}-\frac{m_2^2}{2p_2}
 \right)L+ O(m^4) \nonumber \\
      & = &   \frac{(m_1^2-m_2^2)}{2\overline{p}}L + O(m^4)    
\end{eqnarray}
 in agreement with Eqns(2.15) and (2.16).
The exact formula gives:
\begin{eqnarray}
\phi_{12}^{LAB}(exact) & = & (E_1 \Delta t_1 - E_2 \Delta t_2) - (p_1-p_2)L 
 \nonumber \\
     & = & \left[\frac{E_1}{v_1}-p_1-\frac{E_2}{v_2}+p_2\right]L
 \nonumber \\
    & = & \left[\frac{m_1^2}{p_1} -\frac{m_2^2}{p_2} \right]L
 \nonumber \\
  & = &   \frac{(m_1^2-m_2^2)}{\overline{p}}L + O(m^4)
\end{eqnarray}
  in agreement with Eqns(2.17) and (2.18). It can be seen that:
\begin{equation}
\phi_{12}^{LAB}(exact)-\phi_{12}^{LAB}(\overline{v}) = (E_1 \Delta t_1 - E_2 \Delta t_2)
 -(E_1-E_2)\Delta t =  \frac{(m_1^2-m_2^2)}{2 \overline{p}}L + O(m^4)
\end{equation}     
 Again, the `equal velocity' formula is seen to neglect O($m^2$) contributions that
 double the value of the oscillation phase.
\par The neglect of important O($m^2$) terms when the equal velocity hypothesis
 is made especially transparent on rewriting Eqn(2.21) in the following
 way:
\begin{equation}
\phi_{12}(exact)  =  m_1 \Delta \tau_1- m_2 \Delta \tau_2
  = (m_1-m_2) \Delta \tau_1 + m_2 ( \Delta \tau_1-\Delta \tau_2)        
\end{equation}
 Since 
\begin{equation}
 \Delta \tau_i = \frac{m_i L}{p_i} = \frac{m_i L}{p_0}+O(m^3)
\end{equation}
 Eqn(2.30) can be written as:
\begin{equation}
\phi_{12}(exact)  =  \frac{(m_1-m_2)}{p_0}[m_1+m_2]L+O(m^4)
\end{equation}
 The equal velocity hypothesis now implies:
 \begin{itemize}
 \item[(a)] Neglect of $\Delta \tau_1-\Delta \tau_2$ i.e. of the second 
 term in the square brackets in Eqn(2.32).
  \item[(b)] Making the replacement: $m_1/p_0 \rightarrow \overline{m}/\overline{p}$
   in the remaining term of Eqn(2.32).
 \end{itemize}
 It is now obvious that (a) and (b) neglect (different) O($m^2$) terms;
  a term $\simeq m_2(m_1-m_2)$ for (a) and a term
  $\simeq (m_1-\overline{m})(m_1-m_2) = (m_1-m_2)^2/2$ for (b). The sum
   of these neglected terms is  $\simeq (m_1^2-m_2^2)/2 = \overline{m}(m_1-m_2)$,
  equal to the terms retained by the equal velocity hypothesis.   
  
 \par Thus the essentially universally employed `equal velocity' assumption
  results in the neglect of O($m^2$) terms in the calculation
  of the vacuum oscillation phase. The existence of these terms explains 
  the factor of two difference between the original calculation of the
  oscillation phase by Gribov and Pontecorvo (which is the correct O($m^2$)
  result) and the `standard phase' given by all calculations that make the
  `equal velocity' assumption.

  \par The calculation of the oscillation phase is now repeated assuming, instead
  of equal velocities for the neutrino mass eigenstates, either equal 
  momenta or equal energies. For equal momenta the following relations hold:
\begin{eqnarray}
  L & = &  \Delta t_i(p)v_i(p)  \\
 E_i(p) & = &  \sqrt{p^2+m_i^2}  \\
 v_i(p)  & = &\frac{p}{E_i(p)} \\
\gamma_i(p) & = &\frac{E_i(p)}{m_i}
\end{eqnarray} 
 These yield, for the invariant phase increment:
\begin{equation}
 \Delta \phi_i(p) = m_i \Delta \tau_i(p) = \frac{m_i \Delta t_i(p)}{\gamma_i(p)}
  = \frac{m_i^2 \Delta t_i(p)}{E_i(p)} = \frac{m_i^2 L}{p}
\end{equation} 
 Comparing with Eqn(2.5) it can be seen that each member of the two equations
 differ only by terms of O($m^4$). Thus, at O($m^2$) the Gribov-Pontecorvo result
  (2.21) is obtained for the oscillation phase.
  For the equal energy case:
\begin{eqnarray}
  L & = &  \Delta t_i(E)v_i(E)  \\
 p_i(E) & = &  \sqrt{E^2-m_i^2}  \\
 v_i(E)  & = & \frac{p_i(E)}{E} \\
\gamma_i(E) & = &\frac{E}{m_i}
\end{eqnarray} 
\begin{equation}
 \Delta \phi_i(E) = m_i \Delta \tau_i(E) = \frac{m_i \Delta t_i(E)}{\gamma_i(E)}
  = \frac{m_i^2 \Delta t_i(E)}{E} = \frac{m_i^2 L}{p_i(E)}
\end{equation}
 Again, all terms in last equation differ from the corresponding ones in the exact
 expression (2.5) only by terms of O($m^4$) so that the oscillation phase agrees
 with the value in Eqn(2.21). 
 It is easily seen that this result is also obtained in the case of either equal momenta 
 or equal energies if the oscillation phase is calculated directly in the laboratory
 system as in Eqn(2.28).
 \par It may be shown, in a similar way, that once a constant value of $\Delta t$ is 
 assumed, the standard oscillation phase is obtained independently of whether
 the energy and momentum are calculated on the assumption of exact energy-momentum
 conservation, or whether the equal momentum and different energy or different
  momentum and equal energy hypotheses are used.
\par Thus, the `standard' oscillation phase is obtained only in the case of the
 `equal velocity' hypothesis. In all cases: `equal velocity', `equal momentum'
 and `equal energy' an unphysical assumption is being made (i.e. one that violates
 conservation of energy and momentum). The study just presented shows that in
 the latter two cases the error of the approximation is only of  O($m^4$) in 
 the oscillation phase, and so is negligible at O($m^2$). However, in the former
 case, the neglected terms are of O($m^2$) and so lead to an incorrect result at this 
 order.    
 \par Finally, in this Section it will be found of interest for the subsequent
  discussion of independent temporal or spatial propagation to calculate the
  separate temporal ($T$) and spatial ($S$) parts, in the laboratory system,
   of the Lorentz-invariant oscillation
  phase $\phi_{12}$:
\begin{eqnarray}
\phi_{12}^{LAB}(T) & = & (E_1 \Delta t_1 - E_2 \Delta t_2)
 \nonumber \\
 & = & \left(\frac{E_1^2}{p_1}-\frac{E_2^2}{p_2}\right)L \nonumber \\
 & = & \left(p_1-p_2 +\frac{m_1^2}{p_1}-\frac{m_2^2}{p_2}\right)L \nonumber \\
  & = & \left[1-\frac{(1+R_m^2)}{4}\right]
   \frac{(m_1^2-m_2^2)}{p_0}L + O(m^4)   \\  
\phi_{12}^{LAB}(S) & = & -(p_1-p_2)L = \frac{(1+R_m^2)}{4 p_0}
 (m_1^2-m_2^2)L  + O(m^4)
\end{eqnarray}  
 On adding (2.43) to (2.44) it can be seen that the Lorentz invariant oscillation
 phase originates entirely from the temporal part, the spatial part being 
 exactly cancelled by a term in the temporal one.
   
\SECTION{\bf{Gaussian Wave Packet Propagation }}
 Very often in the literature, following Reference~\cite{GKL}, the plane wave
  propagator of
 Eqn(2.1) has been modified by the introduction of a multiplicative Gaussian
 spatial wave packet: $G(\Delta x, \Delta t, v_i, \sigma_x)$:
\begin{equation}
P_{WP}(\Delta x, \Delta t, m_i, v_i, \sigma_x) = G(\Delta x, \Delta t, v_i, \sigma_x)
P(\Delta x, \Delta t, m_i)
\end{equation}
where $P(\Delta x, \Delta t, m_i)$ is given in Eqn(2.1) and :
\begin{equation}
G(\Delta x, \Delta t, v_i, \sigma_x) = (\sqrt{2 \pi} \sigma_x)^{-\frac{1}{2}}
\exp\left[-\left(\frac{\Delta x - v_i \Delta t}{2 \sigma_x}\right)^2\right]
\end{equation}
 In this formulation the intervals $\Delta x$ and $\Delta t$ are assumed to be the same
 for all mass eigenstates, so that the equal velocity hypothesis is tacitly introduced.
 To calculate the oscillation phase, $\phi_{12}^{WP}$ an integration over the
  transit time in the range $-\infty < \Delta t < \infty$ is performed, leading to
  the result ~\cite{GKL}:
\begin{equation}
 \phi_{12}^{WP} = \left[\frac{E_1-E_2}{\tilde{v}}-(p_1-p_2)\right]L
\end{equation}
 where 
\begin{equation}
 \tilde{v} = \frac{v_1^2+v_2^2}{v_1+v_2} = 1-\frac{1}{2}\left(\frac{m_1^2}{p_1^2}
+ \frac{m_2^2}{p_2^2}\right) + O(m^4)
\end{equation}
 It may be noted that the effective average velocity $\tilde{v}$ is {\it lower} than that
 of either mass eigenstate. This is presumably due to the contribution of the
 unphysical\footnote{Note that the Gaussian variation of $\Delta t$, for fixed source
 detector distance $\Delta t = L$, assumed in this calculation, is at variance with the known
 space-time structure of the decay and detection events, i.e. an exponential
  distribution of decay times followed by a mass dependent propagation time $\Delta t_i$.}
 negative region in the integration over $\Delta t$. 
Using the relation:
\begin{equation}
E_i = p_i+\frac{m_i^2}{2p_i} + O(m^4)
\end{equation}
Eqn(3.3) may be written:
\begin{eqnarray}
 \phi_{12}^{WP} & = &\left\{ \left[(p_1-p_2)+\frac{1}{2}\left(\frac{m_1^2}{p_1}
- \frac{m_2^2}{p_2}\right)\right]\left[1+\frac{1}{2}\left(\frac{m_1^2}{p_1^2}
+ \frac{m_2^2}{p_2^2}\right)\right]-(p_1-p_2)\right\} + O(m^4) \nonumber \\
 & = & \left[\frac{1}{2}\left(\frac{m_1^2}{p_1}
- \frac{m_2^2}{p_2}\right)+\frac{(p_1-p_2)}{2}\left(\frac{m_1^2}{p_1^2}
+ \frac{m_2^2}{p_2^2}\right)\right]L + O(m^4) \nonumber  \\
 & = & \frac{(m_1^2-m_2^2)}{2\overline{p}}L  + O(m^4)
\end{eqnarray}
 where Eqn(2.6) has been used to write $p_i$ in terms of $p_0$ and $m_i$. 
 The  `wave-packeted' propagator thus gives the same oscillation phase at
 O($m^2$) as the unmodified invariant plane wave propagator when the same
 equal velocity hypothesis is made. In both cases important O($m^2$) terms
 are neglected in comparison with the exact calculation. 
\par In a more elaborate recent calculation~\cite{Giunti02} where wave packets are
 associated not only with the neutrinos, but also with all particles participating
 in the production and decay processes, a formula for the phase increment identical
 to (3.1) was obtained except that the width of the spatial wavepacket $\sigma_x$
 is replaced by a parameter $\eta$ that depends on the widths of the wavepackets
 of all participating particles. Since the oscillation phase
 (3.3) does not depend on this parameter the same `standard' result is obtained 
 as in Reference~\cite{GKL}. The result obtained with different mean velocities,
  $\overline{v}$ given by Eqns(2.9)-(2.12) and $\tilde{v}$ given by Eqn(3.4) are the same. 
   Indeed, it is clear from the derivation of (3.6) from (3.3) that a mean velocity
  $\langle v \rangle$ defined by any formula of the type:
\begin{equation}
 \langle v \rangle = 1 - \frac{\alpha m_1^2 +
  \beta m_1 m_2+\gamma m_2^2}{\overline{p}^2} + O(m^4)
\end{equation}
 where $\alpha$, $\beta$ and $\gamma$ are arbitary coefficients of order unity,
 will yield the result (3.6). However, in the derivation of the same formula in Eqns(2.15)
 ,(2.16) this result requires that $\overline{m} = (m_1+m_2)/2$. If instead the value
  $\tilde{m} = \sqrt{m_1^2+m_2^2}$, suggested by Eqn(3.4), is used, then a different
  result is obtained:
\begin{equation}
 \phi_{12}^{I}(\tilde{v}) =  \phi_{12}^{II}(\tilde{v}) =
 \frac{(m_1-m_2)(\sqrt{m_1^2+m_2^2})L}{\overline{p}} + O(m^4) 
\end{equation}
 Thus, unlike in the case of the invariant plane wave, the wave-packet treatment
 gives inconsistent results  for the oscillation phase in the neutrino centre-of-mass
 and laboratory systems.

\SECTION{\bf{Independent Temporal or Spatial Propagation }}
  Many descriptions of neutrino oscillations (especially concise presentations
 in review articles concerned mainly with the description of experimental
 results, or theoretical models of neutrino masses) do not use a Lorentz invariant
 description but consider instead independent temporal (TE) or spatial (SE) evolution
 of the wavefunctions  of the neutrino mass eigenstates. It is in this way that 
 the standard formula for the oscillation phase of Eqn(1.5)
 was first derived in Reference~\cite{FritMink}. A simple Schr\"{o}dinger time 
 evolution is used giving the temporal propagator:
\begin{equation}
 P_{TE}(\Delta t, m_i) = P_0 \exp\left[-iE_i \Delta t\right] 
\end{equation}
 or 
\begin{equation}
 \Delta \phi_i(TE) = E_i \Delta t
\end{equation} 
To relate the time interval $\Delta t$ (assumed to be the same for both neutrino
 mass eigenstates) to the experimentally measured source-detector separation $L$,
 it is further assumed that the neutrinos are ultra-relativistic and so move at the
 speed of light, i.e. $\Delta t = L$. To derive the oscillation phase the
  eigenstates are then assumed to have equal momentum, $p_{\nu}$, but different energies.
  As first pointed out in Reference~\cite{Win} this assumption is in 
  contradiction with energy-momentum conservation in the decay process. 
  This is easily seen by inspection of Eqns(2.6) and (2.7) above. Indeed, the
 violation of energy-momentum conservation is an O($m^2$) effect. Still,
  making this assumption:
\begin{equation}
 E_i = p_{\nu} +\frac{m_i^2}{2p_{\nu}} + O(m^4) 
\end{equation}  
 so that Eqn(4.2) gives:
\begin{equation}
 \phi_{12}(TE) =  \Delta \phi_1(TE)-\Delta \phi_2(TE) =
 \frac{(m_1^2-m_2^2)L}{2p_{\nu}} + O(m^4) 
\end{equation} 
 This calculation is wrong, firstly because it does not respect Lorentz 
 invariance and secondly because the kinematical approximations made
  (neutrinos of equal momenta moving at the speed of light) neglect important
  O($m^2$) terms, as can be seen by comparing the exact expression of the
  temporal part $\phi_{12}^{LAB}(T)$ in the laboratory system of the Lorentz invariant 
   oscillation phase in Eqn(2.43) with Eqn(4.4). 
  \par Returning now to the question of the compatibility of Eqn(4.1) with
  special relativity, such a formula is a reasonably good approximation
  to the exact, Lorentz-invariant, result in the 
  non-relativistic (NR) limit where $ p_i \ll m_i$. In
  this case:
\begin{equation}
\Delta \phi_i^{LAB}(NR) = E_i \Delta t_i = \frac{m_i^2 L}{p_i}\left(1+\frac{p_i^2}
 {m_i^2}\right)
\end{equation}
 so that, up to a fractional correction $p_i^2/m_i^2$, the non-relativistic result
 is the same as the exact one (2.5), and the same result, Eqn(2.21), is obtained
 for $\phi_{12}$. The different value obtained in Eqn(4.4) evidently is a
 consequence of the `equal time' assumption $\Delta t_1 = \Delta t_2 =
  \Delta t$, which results in the neglect of important O($m^2$) terms. It should
  be stressed that the use of Eqns(4.4) or (4.5) to describe ultra-relativistic
  neutrinos is wrong in just the same way that the use of the non-relativistic
  formula for the total energy of a particle, $ E = m + T$, where $T=m v^2/2$,
  is wrong when $v \simeq 1$.
  \par Other authors~\cite{Lipkin1,GroLip,Lipkin2} motivated by the correct
   observation that flavour oscillation experiments actually measure quantum
   interference effects as a function of space, not time, have proposed to
   describe neutrino oscillations in terms of only a spatial evolution of
   the wavefunction:
\begin{equation}
 P_{SE}= P_0 \exp [ip_i L]
\end{equation}
  or 
\begin{equation}
 \Delta \phi_i (SE) = -p_i L
\end{equation}
 Now, making the additional assumption (incompatible with energy-momentum
 conservation in the decay process) of equal energies, $E_{\nu}$, so that:
\begin{equation}
 p_i = E_{\nu}-\frac{m_i^2}{2E_{\nu}} + O(m^4) 
\end{equation}
 Eqn(4.7) gives:
\begin{eqnarray}
  \phi_{12} (SE) &  = & \Delta \phi_1 (SE) - \Delta \phi_2 (SE)
  = \frac{(m_1^2-m_2^2)L}{2E_{\nu}}+ O(m^4) \nonumber \\
 & = & \frac{(m_1^2-m_2^2)L}{2p_{\nu}}+ O(m^4)
\end{eqnarray}
 in agreement with Eqn(4.4). As in the case of independent time evolution,
 Eqn(4.7) is wrong, firstly, because it does not repect Lorentz invariance,
 and secondly kinematical assumptions are made (equal energies
 and different momenta) leading to a result differing by O($m^2$)
 terms from the spatial part of the exact invariant result, as can be seen
 by comparing Eqns(4.9) and (2.44). Unlike for the case of time evolution,
 there is no kinematical domain in which an equation
 similar to Eqn(4.7) is a good approximation.
 It is tantamount to performing a kinematical calculation with 4-vectors,
 the temporal components of which have all been arbitarily set to zero.
 In all cases it yields a very bad approximation. Actually, 
 to derive Eqn(4.6)~\cite{Lipkin1,GroLip,Lipkin2} the Lorentz invariant
 plane wave was written down and equal velocities, i.e. $\Delta t_1
  = \Delta t_2 =  \Delta t$, as well as equal energies were assumed
 for both mass eigenstates so that the temporal contributions cancel
 exactly in the interference phase. This is equivalent to assuming the
 purely spatial propagation of Eqn(4.6). The physical arguments given in
   References~\cite{Lipkin1,GroLip,Lipkin2} to justify the equal energy
  hypothesis are further critically examined in Section 6 below.

 \par It should be noted that both the TE and SE hypotheses contain a further 
  internal contradiction, different from that of the equal velocity hypothesis
  discussed in Sections 3 and 4 above. In the latter, the correct 
  values of $E_i$ and $p_i$, as given by energy-momentum conservation,
  are assigned in the plane wave propagator, only the geometrically inconsistent
  values $\Delta t_1 = \Delta t_2 = \Delta t$ are assumed. For the TE and SE hypotheses, not only
  are the values of $\Delta t_i$ inconsistent with the relation
   $\Delta t_i = L/v_i= L E_i/p_i$, but energy-momentum
  conservation is also violated by the `equal momentum' 
  or `equal energy' hypotheses. The latter assumptions, however, only generate 
  shifts of O($m^4$) in the oscillation phase as compared to the result obtained
  assuming exact energy-momentum conservation.

\SECTION{\bf{The Quantum Mechanics of Flavour Oscillations}}

  In this Section, a general discussion of the underlying principles
 of flavour oscillation calculations in the Feynman path amplitude
 formulation of QM is first given before deriving
 the oscillation phase for a simple two-flavour problem with the
 source particle at rest. More refined calculations performed in
 a similar manner may be found in Reference~\cite{JHF1}.
 \par After the pioneering conceptual work of Planck, Einstein and 
  Bohr, QM was formulated in 
  several independent ways; most importantly, this was done by 
  Heisenberg, Schr\"{o}dinger and Feynman. The Heisenberg (Matrix Mechanics)
  and Schr\"{o}dinger (Wave Mechanics) formulations originally competed for
  the description of atomic physics. However, the greater similarity with
  classical physics, the less abstract nature and superior calculational
  power of the Wave Mechanics approach resulted in this soon being adopted
   as the standard one, both in atomic physics research and in text books
  on non-relativistic QM.  Wave Mechanics is particularly
  well adapted to subjects such as atomic, molecular and nuclear physics,
  where the solution of bound state problems plays an important role;
  a domain that may be called `Quantum Statics'. In the present
  writer's opinion it is, however, less well suited to `Quantum Dynamics'
  where predictions concerning measurements of events at different
  space-time points are required. Flavour oscillation experiments are of
  just this latter type. The formulation of QM most naturally adapted 
  to such problems is the Feynman Path Integral.
  \par The story of this approach is really that of the re-introduction
  of space-time into microscopic physics after the quantum revolution
  of the early 20th Century. Bohr declared in his `Copenhagen
  Interpretation' talk at Lake Como~\cite{Bohr} that space and
  time were outmoded classical categories in the quantum world
   \footnote{In Reference~\cite{Bohr}, Bohr wrote: `Notwithstanding the difficulties
   which, hence, are involved in the formulation of the quantum
   theory, it seems, as we shall see, that its essence may be expressed in
   the so-called quantum postulate which attributes to any atomic 
   process an essential discontinuity, or rather individuality,
   completely foreign to the classical theories and symbolised by
   Planck's quantum of action. This postulate implies a renunciation
   as regards the causal space-time co-ordination of physical 
   processes.'}.
  The first important step in the direction of the rehabilitation of
  space-time was Dirac's seminal paper on the Lagrangian in
  QM~\cite{Dirac}, that was the basis Feynman's formulation
  of the principles of QM specifically in terms of probability 
  amplitudes corresponding to particles moving in
   space-time~\cite{Feyn2,Feyn3}. The essence of this
  approach is contained in a single formula, the importance of which,
  for a fundamental understanding of QM, had already been stressed
  by Heisenberg in 1929~\cite{Heisenberg}:
  \begin{equation}
 P_{FI} = \sum_m \sum_l \left|\sum_{k_n}... \sum_{k_2}\sum_{k_1}\langle f_m| k_n \rangle
...\langle k_2| k_1 \rangle\langle k_1| i_l \rangle \right|^2  
 \end{equation}
  This equation gives the probability of transitions between a group $I = \sum_{l} i_l$ of
  initial states and a group $F = \sum_{m} f_m$ of final states.
 The use of this formula to describe a quantum experiment requires 
  knowledge of both the initial (prepared) and final (measured) states.
  Indeed, the corresponding quantum experiment is defined by the experimenter's
   choice of these states. Only when they are specified is a meaningful 
   comparison of theory with experiment possible. Since typical experimental
  conditions do not permit to either prepare or measure a single quantum
  state the (incoherent) sums over $l$ and $m$ in (5.1) are necessary. These
  sums are defined by purely experimental criteria (detector sizes or resolution)
  and are unrelated to microscopic, quantum level, parameters.
  The states $|k_n \rangle... |k_2 \rangle, |k_1 \rangle$ (of which, in the detailed
   description of any actual space-time quantum experiment,
   there are, actually, a multiple
   infinity) represent {\it unobserved} intermediate states. In the case of 
    particle trajectories in space-time, described by space-time coordinates, $x_i$,
  of interest for the present discussion, the phase to be assigned to any given
  path amplitude, say,
  $\langle f_m| x_n\rangle... \langle x_2|x_1\rangle\langle x_1|i_l\rangle$ is given
   by the corresponding classical action divided by $\hbar$.
    That is, the integral of the 
   classical Lagrangian (divided by $\hbar$) along the classical particle
   trajectory corresponding to the sequence of space-time
  points $x_1$, $x_2$,...$x_n$ where $t_1<t_2...<t_n$. An important feature of Eqn(5.1) is  that
  interference occurs only between amplitudes corresponding to {\it different
  unobserved intermediate states} never between different
  initial and final states. 
  \par For the flavour oscillation problem, the general path integral formula
       (5.1) may be simplified considerably. The initial and final states correspond
   to the space-time positions~\footnote{In the following, quantum states
    will be specified in `space-time' rather than in `configuration space' as
    is conventionally done. The latter definition is convenient when describing, say, 
    the spatial wavefunction of a bound state, such as the hydrogen atom, 
    which requires 6 spatial coordinates for its complete specification. 
    However, flavour oscillation experiments relate events at different
    space-time, not configuration space, positions.} of the source particle 
    and the detection event. As the particles propagate over macroscopic distances,
    in free space, they follow classical (straight line) trajectories. The
    corresponding Feynman path integral then reduces to a Green function, or
    propagator, that is the Fourier transform of the well-known momentum
    space propagator of a free particle in quantum field theory.
    The explicit, general, expression for
    the invariant space-time propagator of a fermion was given in one of Feynman's
    early papers on QED~\cite{Feyn1}. More recently, it has been rederived
    directly, from the covariant path integral, for an arbitary massive particle,
    by Mohanty~\cite{Moh1}. Apart from solid angle correction factors, 
    that are not relevant to the conventional 1-dimensional discussion of
    flavour oscillations, the propagator for on-shell particles, or
    for virtual particles over macroscopic time-like invariant
    intervals such that $c\Delta \tau \gg 1/m_P$ (where $m_P$ is the pole
    mass of the particle) is given, for a stable particle\footnote{
     An unstable particle is described by adding a negative imaginary term
     to the pole mass: $m_P \rightarrow m_P-i\Gamma/2$.}, up to constant factors,
     by the simple
    expression: $\exp(-im_P \Delta \tau)$. This means that it is mathematically
    identical to the `Lorentz invariant plane wave' of Eqn(2.1). However, its
    physical meaning: a Green function that gives the amplitude for a 
    particle to be found at a given space-time point, when, at some prior
    time, the particle was situated at another, well defined, spatial
    position, is quite different to that of the `plane waves' (energy
    momentum eigenstates) of conventional Schr\"{o}dinger-Born
    Wave Mechanics. The latter, interpreted as wavefunctions according to
   the Born prescription, are not square integrable, and so are devoid
   of any information concerning the position of the particle. In the
    case of flavour oscillation experiments, on the other hand, the 
   experimental knowledge that can be obtained concerning the position
   of the propagating particles, once they are created by the decay process,
   is essentially classical, i.e., unaffected by uncertainties of a purely
    quantum mechanical nature. The meaning of this remark will be clarified 
    by a concrete example to be given below. In fact `quantum uncertainty' 
    as manifested in the Heisenberg Uncertainty Relations results
    only in the uncertainty in the time at which the propagating particle
    is produced by the unstable source particle. The 
    associated spread in the physical mass of the source particle, and of 
    any unstable recoil particles, then produces, via energy-momentum 
    conservation in the decay process, momentum or velocity smearing of
    the neutrino mass eigenstates.  
    \par To take a specific example, consider the decay of a pion
     at rest: $\pi^+ \rightarrow \mu^+ \nu_i$. As the pion is unstable,
     it has, in general, a physical mass $W_{\pi}$ that is different from
     the most likely physical mass, the pole mass, $m_{\pi}$. Similarly,
     the physical mass $W_{\mu}$  is different from the pole mass
     $m_{\mu}$. The Breit-Wigner widths, $\Gamma$, of the physical 
     mass distributions are related to the particle mean lifetimes,
     $\tau$, by the energy-time Uncertainty Relation $\Gamma \tau = 1$.
     Since the initial state of the path amplitudes, that of the
     source pion, is the same in all path amplitudes, the kinematical
     effects of the non-equality of $W_{\pi}$ and $m_{\pi}$ will be
     the same for all path amplitudes, i.e. the resultant velocity 
     smearing of the neutrinos is incoherent. On the other hand,
     since the recoil muon is unobserved, its physical mass may be
     different in the paths corresponding to different mass 
     eigenstates. The corresponding Breit-Wigner amplitudes will
    then create a coherent velocity (or momentum) wave packet
    for each eigenstate. This is a simple application of Eqn(5.1).
    The spatial trajectory, for fixed momentum, $p$, yielding  the
     path amplitude:
\begin{equation}
  PA = \langle f_m| x_n, p \rangle... \langle x_2, p|x_1, p\rangle
 \langle x_1, p|i_l\rangle
\end{equation}
is replaced by the sum of path amplitudes (path integral):
\begin{equation}
  PI = \sum_{q} PA_q =  \sum_{q} \langle f_m| x_n, p_q \rangle... \langle x_2, p_q|x_1,
 p_q\rangle\langle x_1, p_q|i_l\rangle
\end{equation}
 As shown in Reference~\cite{JHF1}, the effect of such wave-packets on the oscillation
 phenomenon is minute, in particular for the case of pion decay, as well as for neutrino
  oscillation experiments in general. As discussed below, it may, however,
   become more important for 
  quark flavour oscillations where the spread in the physical masses of the
  propagating particles is of the same order as their mass difference.
\par All the relevant constraints imposed by the Heisenberg Uncertainty Relations
  have been taken into account in the above discussion. In particular, as will be
  considered in more detail in the following Section, there is no `spatial wave
  packet' associated with the neutrinos, and the possibility, (or not) of interference
  between the
  path amplitudes does not depend upon the existence (or not) of such a hypothetical
  spatial wave packet. The momentum wave packet referred to above is not related,
  by a Fourier transform, to any spatial wave packet with simple physical
  properties. The muon is an unstable particle with
  an exponential decay law and mean lifetime $\tau_{\mu}$. The Fourier 
  transform of this exponential yields, in energy space, a Breit-Wigner amplitude
  with width parameter $\Gamma_{\mu} = 1/\tau_{\mu}$ that describes the distribution
   of the physical mass, $W_{\mu}$, of the muon. Using energy-momentum
  conservation, the corresponding spread in $W_{\mu}$ generates 
  a spread in the momentum and energy of the neutrino. This spread, weighted by the Breit-Wigner
  amplitude, generates the momentum wave packet referred to previously. Taking the
  Fourier transform of the Breit-Wigner amplitude will give back the original
  exponential decay law, not a spatial wavepacket.
  \par The irrelevance of the
  momentum-space Uncertainty Relation to the problem under discussion is 
  demonstrated by applying it to the neutrino momentum wave packet in pion
  decay just discussed. With $\Delta p_{\nu} = m_{\mu} \Gamma_{\mu}/m_{\pi}$~\cite{JHF1}
  the corresponding spatial uncertainty is $\Delta x_{\nu} = 1.27$ km. Does this
  give a limit on the possible knowledge of the spatial position of the neutrino?
  By measuring the time of decay of the pion through detection of the
  decay muon, with the easily obtained precision of $10^{-10}$ sec, 
  the subsequent position\footnote{It's direction can be determined by simultaneous
  measurement of the direction of the decay muon} of the neutrino is known
  with an uncertainty of $c \times 10^{-10} = 3$ cm, since the speed of
  light is precisely known. The known uncertainty in the position of the neutrino
  is then a factor $4 \times 10^4$ smaller than $\Delta x_{\nu}$ as calculated from
  the Uncertainty Relation. The latter evidently has no relevance to the
  actual or possible knowledge of the spatial position of the neutrino.
  Except for the spread in decay times of the parent pion (described by
  the energy-time Uncertainty Relation) there is then no `quantum 
  uncertainty' on the position of the neutrino. Once created, its
   position can be known with classical precision.
  \par The final state in Eqn(5.1), in its application to flavour oscillations,
   is the spatial wavefunction of the detection event. At the microscopic
   level, this can be considered as localised to within an uncertainty given
   by the spatial wavefunction of the target particle, which is of atomic 
   dimensions (say $10^{-8}-10^{-7}$ cm). This, however, does not reflect
   the experimental knowledge of
   the position of the detection event, that is limited, by considerations
   of experimental resolution, to the group of states $F$ in Eqn(5.1). In 
   a typical experiment, the experimental resolution might be of the order
   of a centimetre,  seven or eight orders of magnitude larger than the microscopic
   localisation of the target particle. Thus, to a very good approximation,
   the states $|f_m\rangle$ can be assumed to be spatial eigenstates. 
   The correction for the actual distribution of detection events is then
   included by performing a suitable average over the oscillation
   probability, as the contribution of different final states is incoherent,
   requiring classical addition of probabilities. An exactly similar
   argument applies to the initial states $|i_l\rangle$ corresponding to
   the spatial position of the source particle at some well defined time.
   These states also can be assumed, for calculational convenience, to be 
   spatial eigenstates. It is important to stress that this
   {\it does not imply} that the momentum uncertainty of the source particle
   is infinite (inconsistent with the assumption that it is rest). This
   is a gross misinterpretation of the meaning of the momentum-space
    Uncertainty Relation.
   Consider the case that the source particle is bound in an atom. The actual
   quantum positional uncertainty, at the time chosen to define the initial state,
   could be included by writing its spatial wavefunction explicitly in the
   path amplitude. However this distance is so much smaller than the
   experimenter's actual knowledge of the position of the source, which is
   all that matters to accurately evaluate $P_{FI}$, that it is evident that
   the same result will be obtained if, for simplicity, a spatial eigenstate
   is assumed for the source particle. In an analagous way, energy-momentum
    eigenstates are conventionally assumed for all initial and final state
    particles in the momentum space calculatation of invariant amplitudes in
    quantum field theory. This in no way affects the correctness of the 
    results obtained. A naive application of the momentum-space
    Uncertainty Relation, as done above for the path amplitude calculation,
    to this case would imply an infinite uncertainty on the space-time
    position of the scattering event or decaying particle. Actually the
    space-time parts of the wavefunctions of the incoming and outgoing
    particles play no role in a momentum space calculation except for imposing
    the constraints of exact energy and momentum conservation. If this is
    assumed {\it ab initio} the space-time parts of the wavefunctions can be
     omitted completely without changing the results of the calculations.
    \par  Of course, in any actual experiment, the source particle 
     is not at rest. If it is 
    negatively charged and forms an atomic bound state it will have some
     Fermi motion. If it is positively charged, and remains unbound, it will
     be subject to random thermal motion. Corrections for these effects
     can be included in the calculation of the oscillation frequency by a
     suitable averaging procedure. Since they affect only the definition
     of the group $I$ of initial states, these corrections are incoherent,
      and, just as in the case of the positions of the source particle and
     detection event, the averaging must be done at the level of probabilities,
     not amplitudes. In fact averaging the source momentum distribution at the
     amplitude, rather than probability, level will result in the generation
     of a spurious `momentum wavepacket' with parameters characteristic 
     of the initial momentum {\it wavefunction}. As discussed in the 
     following Section, just such a mistake has been made in many papers
     on the QM of neutrino oscillations in the literature. 
     \par Before defining the intermediate states $|k_i \rangle$ in Eqn(5.1)
     for the case of lepton flavour oscillations, it is important to consider the
     nature of the decay process in which the neutrino mass eigenstates
     $|\nu_1\rangle$ and $|\nu_2\rangle$ are produced, as already mentioned
     in Section 1. In the Standard 
     Electroweak Model, these states are created by diagrams containing real 
     or virtual W bosons. In the case that neutrinos are massive, and the
     different eigenstates have different masses, there is a non-diagonal,
     unitary, flavour/mass mixing matrix, the
     MNS~\cite{MNS} matrix, that is strictly analgous to the better-known
     CKM~\cite{CKM} quark flavour/mass mixing matrix. 
     The existence of such a non-diagonal matrix implies that lepton (or generation)
     number is not
     conserved in transitions in the lepton sector just as generation number
     is not conserved in the quark sector because of the non-diagonal
     nature of the CKM matrix.
     For example, the decays:
     \begin{equation}
      ~~~~~~~~~\pi^+ \rightarrow (\rm{W}^+)^{\ast} \rightarrow \mu^+ \nu_1,~~~~~~~
     \pi^+ \rightarrow (\rm{W}^+)^{\ast} \rightarrow \mu^+ \nu_2
     \end{equation}
      where $ (\rm{W}^+)^{\ast}$ denotes a virtual W$^+$ boson, are strictly analagous to the quark sector processes, where an on-shell W$^+$ boson decays into quarks:
     \begin{equation} 
      ~~~~~~~~~\rm{W}^+ \rightarrow \rm{c} \overline{\rm{s}},~~~~~~~
     \rm{W}^+ \rightarrow \rm{c} \overline{\rm{d}}
     \end{equation}
      in which the first (second) transitions in Eqn(5.5) are Cabibbo allowed 
      (suppressed).
      the neutrino mass/flavour mixing angle $\theta$ in Table 1 plays the
      same role in (5.4) as the Cabibbo angle $\theta_c$ in (5.5). Note that the
       $\mu^+$ in (5.4) and the c quark in (5.5) are both second generation
       particles, whereas the decay antiquarks are either first or second 
       generation. This suggests that the neutrino mass eigenstates should also 
       be associated, as are the quark mass eigenstates,
       to different generations. Just as the W decays in (5.5)
       are {\it independent physical processes}, so also are the pion decays in
       (5.4). This has the important consequence, for the following discussion,
        that $|\nu_1\rangle$ and  $|\nu_2\rangle$ may be produced at 
        {\it different times} in the corresponding path amplitudes. As will be
        seen, this leads to an important new contribution to the oscillation
        phase. The incoherent nature, as in (5.4) above, of decays into different
        mass eigenstates has been previously pointed out by Shrock~\cite{Shrock1,Shrock2}.  
        \par What has been assumed until now in all discussions of the QM
         of neutrino oscillations in the literature is that is that lepton number
          is conserved
         in pion decay, so that the unique physical process is
         $\pi^+ \rightarrow \mu^+ \nu_{\mu}$. Thus the `flavour momentum
         eigenstate' $| \nu_{\mu}\rangle$ is assumed to be produced at a fixed
         time. It is then rewritten, at this time, in terms of
          $|\nu_1\rangle$ and  $|\nu_2\rangle$ using the mixing amplitudes
         given in Table 1. These mass
         eigenstates subsequently evolve in time according to the different
         propagators discussed in Sections 2, 3 and 4 above. The present writer's
       opinion is that this procedure is wrong. If neutrinos of different
       generations have different masses, lepton number {\it cannot} be 
       conserved at charged current weak vertices, just as, in the quark sector,
       generation number is not conserved at such vertices. The 
       non-existence of a consistent theoretical description
       of `flavour momentum eigenstates' has been previously
       stressed in the literature~\cite{GKL1}. Once the incoherent nature of,
       for example, the two processes in (5.4), is
       recognised, it will be seen that the conceptual problems that have 
       beset the treatment of the QM of neutrino osillations in the past,
       as evidenced by the many kinematical and geometrical inconsistencies
       discussed above, all disappear. The neutrino oscillation 
       phenomenon occurs even when the constraints of of both exact
        energy-momentum conservation and space-time geometry are imposed. The
        latter is simply the condition that the source particle is defined and the
       detection event observed, each at a unique space time point, and that
        both the source particle and the neutrinos follow classical space-time
       trajectories that, together, link these two points.
       \par The detailed application of Eqn(5.1) to the derivation of the 
        oscillation probability for a simple two-flavour problem, with the
        source at rest, will now be discussed. A specific example is given by
        `$\nu_{\mu} \rightarrow \nu_e$'\footnote{`$\nu_{\mu} \rightarrow \nu_e$' 
         is only a shorthand notation to indicate that a muon is involved in the
         production process and that a electron is observed in the detection event.
         This notation is used, without quotes below, as is conventional. It should,
          however, not be forgotten that $\nu_{\mu}$ and $\nu_e$ do not exist as 
          physical particles if neutrinos are massive.}  
         oscillations following the decay, at rest,
        of a $\pi^+$. The initial state corresponds to the undecayed source 
        particle, assumed to be at rest (or to have, in any case, a very small
        and random Fermi or thermal motion), at some arbitary time $t_0$. At later
        times $t_1$ or $t_2$ the source particle may decay into the eigenstates
         $|\nu_1\rangle$ or  $|\nu_2\rangle$ respectively. These two possiblities
        are, of course, in classical physics, mutually exclusive. In QM however,
        as the two different decay modes are indistinguishable, when only the
        detection event is observed, the corresponding amplitudes, not 
        probabilities, must be added. The final state corresponds to the detection
        event, which occurs at time $t_D$ at distance $L$ from the source, and which 
        may be produced by an interaction of either of the (indistinguishable)
        neutrino mass eigenstates with a particle of the detector.
         Suppose now that $m_1 > m_2$. Since either 
        neutrino mass eigenstate may create the detection event, and since
        $v_2 > v_1$, it follows that $t_2 > t_1$. 
        \par The different quantum states in Eqn(5.1) may now be identified.   
          The initial state is $|S_l,t_0\rangle$ where $S$ denotes the source 
         particle. The label $l$ corresponds to different source positions in
         the experimental apparatus, as well as, possibly, small random momenta
         of the source particle. The source wavefunction then propagates in time (but not 
         in space) over the interval: $t_0 < t < t_i$. This evolution is described,
         in the case of a source particle of mass $m_S$ and decay width $\Gamma_S$,
       by the Feynman invariant space-time propagator~\cite{Feyn1,Moh1}:
  \begin{equation}
  \langle S_l, t_i| S_l, t_0\rangle = \exp \left[(-im_S-\frac{\Gamma_S}{2})
  (\tau_i-\tau_0)\right]
   \end{equation}
\[ ~~~~~~~~~~~~~(\rm{source~particle~of~mass~}m_S)\]
 where $\tau_i$ and $\tau_0$ are the proper times in the rest frame of S
  corresponding to the laboratory times $t_i$ and $t_0$. For the case of a
  source which is a $\beta$-radioactive nucleus, where the decay process is
   more appropriately described by non-relativistic quantum mechanics,
  the source propagator is:
  \begin{equation}
  \langle S_l, t_i| S_l, t_0\rangle = \exp \left[(-iE_{\beta}-\frac{\Gamma_S}{2})
  (\tau_i-\tau_0)\right]
   \end{equation}
\[~~~~~~~~~~~~~~( \beta-\rm{decay~of~a~nucleus})\]
  where, neglecting the recoil energy of the daughter nucleus, $E_{\beta}$ is
   the total energy release in the $\beta$-decay process.
  The formula (5.7) may be derived in the same way as the analagous one for
  a radiative transition in atomic physics, by the use of perturbation theory
  and the time-dependent Schr\"{o}dinger equation~\cite{Heitler}.
   The appropriate amplitude to describe the $\beta$-decay transition is then
   the matrix element $\langle N_f|T| N_i\rangle$ where $N_i$ and $N_f$ denote the
    parent and daughter nuclei respectively. Since these are in energy eigenstates
    the Schr\"{o}dinger Equation predicts that they evolve in time as:
    $|N_j,t \rangle = |N_j,t_0 \rangle \exp[-iE_j(t-t_0)]$, ($j=i,f$).
    It follows that:
  \begin{eqnarray}
  \langle N_f, t |T| N_i, t\rangle & = & \langle N_f, t_0 |T| N_i, t_0 \rangle
   \exp[-i(E_i-E_f)(t-t_0)] \nonumber \\
    &   & \simeq \langle N_f, t_0 |T| N_i, t_0 \rangle
   \exp[-iE_{\beta}(t-t_0)]
   \end{eqnarray}
  where, in the last member of Eqn(5.8), the recoil energy of the nucleus 
  $N_f$ has been neglected. In the path amplitude formalism, 
  $ \langle N_f, t_0 |T| N_i, t_0 \rangle$ may be considered as a time 
  independent amplitude (the same in all interfering path amplitudes)
  while the exponential factor in (5.8) represents the propagator of the
  parent nucleus. It is assumed, for simplicity, in (5.8) that $t-t_0 \ll 1/\Gamma_S$.
  For a source at rest where $t = \tau$, Eqn(5.8) leads to (5.7). It is interesting
  to note that, if $N_i$ and $N_f$ in (5.8) are replaced, for pion decay, by the quark degrees 
   of freedom, via the correspondences: $|N_i \rangle \simeq |\pi \rangle$ and 
    $|N_f \rangle \simeq | 0 \rangle$, where  $| 0 \rangle$ denotes the vacuum
   (zero energy) state of the quark sector, then the rest frame pion propagator
   is predicted by Eqn(5.8) to be : $\exp[-im_{\pi} \Delta \tau]$, in agreement
   with the Lorentz invariant Feynman space-time propagator of Eqn(5.6) above. 

 \par The source
  particle then decays into a final state containing $\nu_i$ with the
  time-independent amplitude:
  \begin{eqnarray}
   A(i \leftarrow S) & = & \langle i|\alpha \rangle \langle \alpha|T|
   S_l \rangle \nonumber \\
    & = &  \langle i|T| S_l \rangle
  \end{eqnarray}
 In the first line of Eqn(5.9), the neutrino mass/flavour mixing amplitude
  is shown explicitly. However, as explained above, like the elements of the
   CKM matrix,  this amplitude is an intrinsic part of the charged current 
   coupling, so that the amplitudes $A(1 \leftarrow S)$ and $A(2 \leftarrow S)$
   describe independent physical processes. Thus there is no reason
   for the assumption $t_1=t_2$ that has been made in essentially all 
   discussions of the QM of neutrino oscillations in the literature to date.
  \par The third element in the path amplitude is the invariant space-time
    propagator of the neutrino~\cite{Feyn1,Moh1}
  \begin{equation}
  \langle i, x_D, t_D| i, x_i, t_i\rangle = \exp [-im_i(\tau_D-\tau_i)]
   \end{equation}
  \par The fourth and last element in the path amplitude 
  is the time-independent amplitude of the
  detection process:
  \begin{eqnarray}
  A(d \leftarrow i) & = & \langle d_m |T|\beta \rangle 
 \langle \beta | i \rangle \nonumber \\
    & = &  \langle d_m |T| i \rangle
  \end{eqnarray}
 where the label $m$ specifies the spatial position of the detection 
 event, as well as possibly the directions and energies of the particles,
   in the final state of the event, that are detected.
  \par The path amplitudes are then  given by combining (5.6) or (5.7),
  (5.9), (5.10) and (5.11):
  \begin{equation}
  A_i(\beta \leftarrow \alpha) = \langle d_m |T| \beta \rangle
  \langle \beta| i \rangle \langle i, x_D, t_D| i, x_i, t_i\rangle 
  \langle i | \alpha \rangle \langle \alpha |T| S_l \rangle
  \langle S_l, t_i| S_l, t_0\rangle
    \end{equation}
  The phase increment $\Delta \phi_i$ can be read off directly from
  Eqns(5.6) or (5.7) and (5.10):
   \begin{equation}
 \Delta \phi_i = m_i (\tau_D-\tau_i) + E_S(\tau_i-\tau_0) 
     \end{equation}
 where $E_S = m_S$ or $E_{\beta}$ for unstable particle or nuclear
  $\beta$-decay sources respectively.
  Now,
   \begin{eqnarray}
 \tau_i-\tau_0  & = &  \frac{t_i-t_0}{\gamma_S} = t_i-t_0 +O(v_S^2)
  \nonumber \\
   & = &  t_D-t_0 -\frac{L}{v_i} +O(v_S^2)  \nonumber \\
   & = &  t_D-t_0 - L\left[1+\frac{m_i^2}{2 p_0^2}\right] +O(v_S^2)+O(m^4)
  \end{eqnarray}
  so that, using Eqns(2.5) and (5.14), Eqn(5.13) may be written
  (neglecting any random motion of the target) as:
 \begin{equation}
 \Delta \phi_i = \frac{m_i^2}{p_0}\left[1-\frac{E_S}{2p_0}\right]L
   +E_S(t_D-t_0) -E_S L +O(m^4) 
 \end{equation}
 giving for the oscillation phase:
 \begin{equation}
  \phi_{12} = \frac{\Delta m^2}{p_0}\left[1-\frac{E_S}{2p_0}\right]L+O(m^4) 
 \end{equation}
 where
  \begin{equation}
   \Delta m^2 \equiv m_1^2-m_2^2
   \end{equation}
  It can be seen that, in addition to the contribution due to neutrino
  propagation, first given by Gribov and Pontecorvo (the first term on
  the RHS of Eqn(5.16)), there is also an important contribution due to the
  propagation of the source particle during the time interval
  $t_1 < t < t_2$. For example, in the case of pion decay at rest,
  $E_S/2p_0 = m_{\pi}/2 p_0 = 2.34$, so that the oscillation phase
  is 34$\%$ larger than the Gribov and Pontecorvo result and 
  a factor 2.68 times larger than the prediction of the standard
   formula. In order to calculate the oscillation probability,
   from flavour $\alpha$ to flavour $\beta$, at the fixed
   source-detector separation, $L$, the probability corresponding
   to the sum of the amplitudes in Eqn(5.12) must be integrated
   over the detection time $t_D$. For this, it is convenient to
   introduce the times-of-flight $t_1^{fl}$, $t_2^{fl}$ of the
   neutrinos: 
 \begin{equation}
 t_i^{fl} = t_D-t_i = L\left[1+\frac{m_i^2}{2p_0^2}\right]+O(m^4)
   \end{equation}
  then,
\begin{eqnarray}
P_{ \beta  \alpha } & = & \int_{t_{min}}^{\infty}|A_1+A_2|^2dt_D 
 \nonumber \\
  & = & |\langle d_m|T|\beta \rangle \langle \alpha|T| S_l \rangle |^2
  \int_{t_{min}}^{\infty}e^{-\Gamma_S(t_D-t_0)}\left[
   (\langle \beta|1 \rangle \langle 1 | \alpha \rangle)^2 e^{\Gamma_S t_1^{fl}}
  + (\langle \beta|2 \rangle \langle 2 | \alpha \rangle)^2 e^{\Gamma_S t_2^{fl}}\right.
 \nonumber \\
  &   &\left. + 2\langle \beta|1 \rangle \langle 1 | \alpha \rangle
   \langle \beta|2 \rangle \langle 2 | \alpha \rangle e^{\Gamma_S (\frac{t_1^{fl}+t_2^{fl}}
  {2})}\cos\frac{\Delta m^2}{p_0}\left(\frac{E_S}{2 p_0}-1\right)L\right]dt_D
\end{eqnarray}
 $t_{min}$ takes the values $t_0+t_1^{fl}$,  $t_0+t_2^{fl}$ and  $t_0+t_1^{fl}$ for 
  the squared amplitudes for neutrinos of masses $m_1$, $m_2$ and the interference
  term respectively, where it is assumed that $m_1 > m_2$, so that $t_1^{fl}>t_2^{fl}$ 
  The $t_D$ integral cancels the time-of-flight dependence of the squared amplitude
  terms and gives a factor $\exp[-\Gamma_S(t_1^{fl}-t_2^{fl})/2]$ multiplying the
   interference term. Thus, using Eqn(5.18) the following time-averaged oscillation
   probability is obtained:
 \begin{eqnarray}
P_{ \beta \alpha } & = & \frac{|\langle d_m|T|\beta \rangle \langle \alpha|T| S_l \rangle |^2}
  {\Gamma_S}\left[ (\langle \beta|1 \rangle \langle 1 | \alpha \rangle)^2
 +  (\langle \beta|2 \rangle \langle 2 | \alpha \rangle)^2 \right.
 \nonumber \\
  &   & +\left. 2\langle \beta|1 \rangle \langle 1 | \alpha \rangle
   \langle \beta|2 \rangle \langle 2 | \alpha \rangle\exp\left(-\Gamma_S
\frac{\Delta m^2 L}{4 p_0^2}\right)
\cos\frac{\Delta m^2}{p_0}\left(\frac{E_S}{2 p_0}-1\right)L\right]
\end{eqnarray} 
 In all conceivable neutrino oscillation experiments the exponential damping
 of the interference term is very small. Since observation of the oscillation
 requires the argument of the cosine in Eqn(5.20) to be of order unity, i.e.,
 $\Delta m^2/p_0 \simeq 1$, the damping term is typically $\simeq \exp(-\Gamma_S
  /4 p_0)$. In the case pion decay, where $\Gamma_S = 2.5 \times 10^{-14}$ MeV,
  and $p_0 = 29.8$ MeV, the damping factor is $\simeq 1-4.1 \times 10^{-16}$.
  For $\Delta m^2 = (1\rm{eV})^2$, the distance $L$, in pion decay at rest,
  is $1.9 \times 10^{16}$ m or 2.0 light yr for 50 $\%$ damping of the
  interference term.
   \par Two important special cases of Eqn(5.20) are $P_{ e \mu }$, as, for example
    in $\nu_{\mu}\rightarrow \nu_e$ oscillations following $\pi^+$ decay 
    ($\nu_e$ appearence) and $P_{e e}$ as in $\nu_e\rightarrow \nu_e$  following
    $\beta^+$-decay using a nuclear reactor as a source ($\nu_e$ disappearence).
     Neglecting the exponential damping of the interference term, and using
     the flavour/mass mixing amplitudes in Table 1, the following
     predictions are obtained:
 \begin{equation}
P_{e \mu}  =  2\frac{|\langle d_m|T|e \rangle \langle \mu|T| S_l \rangle |^2
  \sin^2\theta \cos^2\theta}
  {\Gamma_S}\left[1-
\cos\frac{\Delta m^2}{p_0}\left(\frac{m_{\pi}}{2 p_0}-1\right)L\right]
\end{equation}
 \begin{equation}  
P_{e e}  =  \frac{|\langle d_m|T|e \rangle \langle e|T| S_l \rangle |^2}
  {\Gamma_S}\left[\sin^4\theta+\cos^4\theta +2\sin^2\theta \cos^2\theta
\cos\frac{\Delta m^2}{p_0}\left(\frac{E_{\beta}}{2 p_0}-1\right)L\right]
\end{equation}
  \par The derivation of Eqn(5.20) has neglected velocity smearing of the 
   neutrinos due to the finite decay widths of the source, and also possibly,
   the recoil particles. These effects have been estimated for the case of pion
   decay (using a Gaussian approximation for the Breit-Wigner amplitudes) in
    Reference~\cite{JHF1}. They are found to be more than ten orders of
    magnitude smaller than the already minute correction term for
    liftime damping discussed above. Corrections to Eqn(5.20) due to finite 
    experimental source and detector sizes , as well as the effect of thermal 
    motion (for a positive pion source) are also calculated in Reference~\cite{JHF1}.
    \par An extended discussion of the application of the Feynman path amplitude
     method to flavour oscillations of neutral kaons and b-mesons will be
     presented elsewhere~\cite{JHF3}. Here, only a few remarks on the salient
     differences with respect to neutrino oscillations will be made.
     \begin{itemize}
     \item[(i)] Because of the very small fractional mass differences between
      different neutral kaon and b-meson mass eigenstates~\cite{PDG}:
   \begin{eqnarray}
   \frac{\Delta m_{\rm{K}}}{m_{\rm{K}^0}} & = & \frac{m(\KL)-m(\KS)}{m_{\rm{K}^0}} = 7.1 \times 10^{-15} \\
   \frac{\Delta m_{\rm{B}}}{m_{\rm{B}^0}} & = & \frac{m(\BH)-m(\BL)}{m_{\rm{B}^0}} = 5.9 \times 10^{-14}
   \end{eqnarray}
   The Lorentz-$\gamma$ factors relating the proper time and laboratory times of the 
    mass eigenstates are then equal to within the
    fractional differences quoted in Eqns(5.23) and (5.24). Thus, unlike for the
    case of neutrinos, the `equal velocity' hypothesis discussed in Section 2
    above, is expected to be a good kinematical approximation.
      \item[(ii)] The unobserved propagating particles $\KS$, $\KL$ and $\BL$,
   $\BH$ are unstable, with decay widths that for $\KS$, $\BL$, and
   $\BH$ are of the same order as the mass
    differences:
  \begin{eqnarray}
  \frac{\Gamma_S}{\Delta m_{\rm{K}}} & = & 2.10 \\
  \frac{\Gamma_L}{\Delta m_{\rm{K}}} & = & 0.037 \\
  \frac{\Gamma_{\BL}}{\Delta m_{\rm{B}}} & = & \frac{\Gamma_{\BH}}{\Delta m_{\rm{B}}} = 1.37 \\
  \end{eqnarray}
   Unlike in the case of neutrino oscillations, where the natural widths of the source
   and recoil particles give only very small variations of velocity, the decay
   widths of the $\KS$, $\BL$ and $\BH$ then produce velocity smearing effects of
   the same order as the velocity differences resulting from the mass differences. Such effects 
   must then be taken properly into account in the calculation of the flavour 
   oscillation probability. However, because $\Gamma_L \ll \Gamma_S$, the natural 
   width of the $\KL$ may, in first approximation, be neglected. Including 
   correctly the kinematical effect of the physical mass, $W$, of the propagating
   particles modifies the `on-shell' expression (2.5) for the phase increment to:
\begin{equation}
 \Delta \phi_i = m_i \Delta \tau_i = \frac{m_i \Delta t_i}{\gamma_i}
  = \frac{m_i W_i \Delta t_i}{E_i} = \frac{m_i W_i L}{p_i}
\end{equation}
 It is important to note that although, for example, the $\KS$ is significantly off-shell
 on the scale of $\Delta m_{\rm{K}}$, the condition: $c\Delta \tau \gg 1/W_S$ is still well
 satisfied in a typical flavour oscillation experiment, so the off-shell $\KS$
 can still be considered to propagate as a classical particle according to Eqn(5.29).
 It may also be remarked that the pole mass, $m_i$, in Eqn(5.29) results from the 
  Fourier transform of the invariant momentum-space propagator, and so is
  constant for each particle species, unlike the variable physical mass $W_i$.
\item[(iii)] Unlike neutrinos which, in terrestial oscillation experiments, are
  produced incoherently as 
  single quantum states from a coherent source (an unstable particle or nucleus)
  neutral kaons may either be produced incoherently, in an incoherent interaction, such as:
  $\pi^-\rm{p} \rightarrow \Lambda(\KL,\KS)$ or $K^-\rm{p} \rightarrow n(\KL,\KS)$, or in
  correlated `entangled' states, from a coherent source, as in: $\phi \rightarrow
  \KL \KS$. To date, B-mesons have usually been produced in entangled states from coherent
  sources, for example  $Z \rightarrow \BL \BH X$ or
  $\Upsilon(4S) \rightarrow \BL \BH$.
 \end{itemize}
 \par Because of the points (ii), (iii) above, the path amplitude analysis of 
   $\KS-\KL$ and $\BL-\BH$ oscillations is more complicated than for the 
   case of neutrino oscillations. The standard procedure, to date, in analysing
   experiments, is to assume equal velocities (that seems, in view of 
   Eqns(5.23) and (5.24) to be reasonable) so that, for example,
   $\Delta \tau_L = \Delta \tau_S =  \Delta \tau$ and use the first member
   of Eqn(5.29), together with the relation: $\Delta \tau = m_{\rm{K}^0}L/p_{\rm{K}^0}$
   to give the formula:
   \begin{equation}
    \Delta \phi_i = \frac{m_i m_{K^0} L}{p_{K^0}} 
   \end{equation} 
    Setting $m_{\rm{K}^0} = (m_L+m_S)/2$ in Eqn(5.30) gives, for the oscillation
    phase: 
   \begin{equation}
    \phi_{LS} =  \frac{\Delta m_{\rm{K}} m_{\rm{K}^0} L}{p_{\rm{K}^0}} = 
     \frac{(m_L-m_S)(m_L+m_S)L}{2 p_{\rm{K}^0}} =  \frac{(m_L^2- m_S^2) L}
    {2 p_{\rm{K}^0}} 
   \end{equation}
   This is the same as the standard formula for neutrino oscillations, also derived on the
   basis of the equal velocity hypothesis. At the time of writing, the author is not
   able to comment on the correctness or otherwise, of Eqn(5.31).
  \par Taking into account exact energy-momentum conservation in the process:
   \newline
      $\pi^-\rm{p} \rightarrow \Lambda(\KL,\KS)$, but assuming all particles to be on-shell,
   the authors of Reference~\cite{SWS1} obtained a different formula to Eqn(5.31),
   as well as predicting correlated spatial oscillations of in the decays
   of the neutral kaons and the lambda.

 \SECTION{\bf{Discussion }}
  There are two essential differences between the work presented in the present
 paper and References~\cite{JHF1,JHF2}, and all previous treatments of the QM of 
 neutrino oscillations in the literature: (i) the realisation that the different
 neutrino mass eigenstates are produced in independent physical processes, not as
 a coherent lepton flavour eigenstate, and (ii), the assumption that, in the covariant
  Feynman path
 amplitude description, the neutrino mass eigenstates follow essentially
  classical space-time trajectories; there is no spatial `wave packet' 
  associated with their propagation. Historically, as will be seen, the failure to 
  notice (i), leading to the universal assumption that the different neutrino mass 
 eigenstates are produced coherently and at the same time, required the introduction
 of a spatial wave packet, since otherwise the neutrino mass eigenstates, produced
 at the same time and propagating with different velocities could not produce the
 detection event at a well-defined space-time point. Only, it was argued, by 
  introducing the spatial `fuzziness' associated with the hypothetical spatial
  wave-packet
  would interference be possible. In the Feynman path description there is no
  such spatial `fuzziness'. The detection event can be produced by the 
  interaction of either (indistinguishable) neutrino mass eigenstate with the
  target particle in the detector. Since the space-time trajectories of the 
  neutrinos are essentially classical, with different velocities, this implies
  different production times for the neutrinos in the alternative interfering
  amplitudes. In view of (i) and the long lifetimes of typical neutrino sources
  different decay times are allowed. This is the fundamental reason for the
  possible occurence of the oscillation phenomenon. The propagator of the source
  particle contributes to the path amplitudes associated with each neutrino
  mass eigenstate, and, as a consequence of the different decay times, gives an 
  important contribution to the interference phase, as demonstrated by the
  calculations presented in Section 5 above. 
  \par The important point (i) above was pointed out many years ago by Shrock
  \cite{Shrock1,Shrock2} but was never, to the author's best knowledge,
    applied before the work reported in this 
   paper and References~\cite{JHF1,JHF2} to the neutrino oscillation problem.
   Indeed, the contrary, incorrect, hypothesis of an initial `lepton flavour 
   eigenstate', that is a coherent superposition of different mass eigenstates,
   has, instead, been universally assumed in the treatment of neutrino oscillations.
   \par A brief historical review of the introduction of the `wave-packet'
    concept into the description of neutrino oscillations will now be made.
    The first descriptions of neutrino oscillations were based on plane-wave
    propagators either (tacitly) covariant in the case of Gribov and Pontecorvo
    ~\cite{GribPont}, or, temporal only, as in the case, for example, of Fritsch
     and Minkowski~\cite{FritMink}  who were the first to derive the standard
     oscillation phase. The wave-packet concept in the context of neutrino 
     oscillations was first introduced by Nussinov~\cite{Nussinov}, together
     with the idea of a `coherence length' for neutrino oscillations. Nussinov
    discussed the effect analagous to collision broadening of atomic spectral
    lines for neutrino sources in the Sun. Both the `wave-packet' and 
     `coherence-length' concepts are related to a classical wave rather than
    a quantum mechanical description of the associated phenomena. A source
    of classical waves of finite duration will evidently produce a wave-train
    of finite length. Fourier analysis of this spatial wave-train will result
     in a spread in the momentum of the associated plane-waves described by 
     a coherent momentum wave-packet. This classical description does not
     match the sequence of spatio-temporal phenomena underlying the line
     broadening effect as described by quantum mechanics. Actually, the
     unstable source produces particles (photons or neutrinos) in a process
     that has a characteristic
     time\footnote{Possible ways to experimentally
     measure such a `process duration' time are discussed in Reference~\cite{DeLeoRot}.}
     much shorter than the mean decay time of
     the source. Interference effects will occur if some process, intitiated 
     by a decay particle, can correspond to different production times of
     the latter. In the case of, say, atomic line broadening as observed in
     the interference-fringe contrast in a Michelson interferometer, this time
     difference results from the different propagation times of the photon
     in the different arms of the interferometer. In the case of
      neutrino oscillations, with a source at rest, it results from different
      times-of-flight of the different mass eigenstates over the same spatial
      distance. In both cases the line broadening effect results from 
       perturbation of the coherent source during the time interval
       between the two emission times corresponding to the quantum
      interference condition. Thus the fundamental physical parameter
     governing the interference damping is the effective lifetime of
     the coherent source, due either to its spontaneous decay rate
     or to collision processes. Nussinov correctly identified this
     parameter, $\tau_{eff}$, as the one controlling the damping
     of the oscillations, but then introduced a hypothetical 
     `spatial wave packet' of length $c\tau_{eff}$. This is 
     actually a classical wave analogue of the series of spatio-temporal
     quantum processes just described. As is well-known from atomic radiative
     transitions the classical wave and QM calculation both lead to a 
     Lorentzian line-shape and so are equivalent if only a description
     of the momentum distribution of the produced particles is required
     \footnote{This is only true as regards the mathematical form of
       the line shape. The width of the distribution predicted by
       the QM calculation depends on the lifetimes of both the initial and
       final states, so there is, in general no direct relation between the
       line width and the length, $c \tau_{eff}$ of the analagous classical
       wave packet.}.
     As shown in the calculations in Section 5, a correct description of interference
     damping requires the spatio-temporal sequence of quantum events to
     be properly taken into account. This is not possible in the classical
     wave description in which a hypothetical spatial wave packet is introduced.
     It must be stressed that in the quantum mechanical description of atomic
     radiative transitions there is no wave packet associated with photon 
    propagation. The Fourier transform of the exponential decay law of
     the initial atom produces a Lorentzian distribution
     in the energy of this atom. When it decays by photon emission,
     energy conservation then produces a smearing in the energy
     (or wavelength)  of the photon that reflects the energy uncertainties
     of both the initial and final atoms. But the photon
     propagates in space-time like a particle with velocity c. There is
     no associated spatial, or momentum, wave packet. Indeed, a photon is
     a particle and not a classical wave. 
      \par Krauss and Wilczek~\cite{KW} who also estimated the effect of 
    collision broadening on solar neutrino sources, although also mentioning
    the wave packet description, understood that  
    neutrino oscillations require different emission times in the 
    interfering amplitudes:
     \par`Since these mass eigenstates propagate with different velocities
         (for fixed energy) the desired interference is between neutrinos
     emitted at {\it different times}.'
     \par Actually, the mass eigenstates do not have fixed energy (or
       momentum) but certainly have different velocities, so that the above
      assertion concerning the interference mechanism underlying neutrino
      oscillations is correct.
  \par The first extended discussion of the QM of neutrino oscillations
   in terms of wave packets was made by Kayser~\cite{Kayser}. In contradiction
   with the conclusions of Shrock published in the previous year~\cite{Shrock1}
   the initial state of the neutrinos was assumed to be a superposition of mass
   eigenstates with definite lepton flavour, and the propagating neutrinos 
   were assumed to have equal momenta and different energies. Thus
   energy-momentum conservation is violated in the neutrino production process.
   Kayser claimed, that in order for the neutrino oscillation phenomenon
   to be possible, the uncertainty in the momentum of the neutrinos must
   satisfy the condition:
   \begin{equation}
     \frac{\Delta p_{\nu}}{ p_{\nu}} \ge \lambda \frac{\Delta m^2}{p_{\nu}^2}
   \end{equation}
   where $\lambda \simeq 10-100$.
   It was then proposed to realise this condition by introducing a hypothetical
    coherent momentum wave packet with a width consistent with the condition
    (6.1). This condition was derived from the momentum-space Uncertainty 
    Relation as follows: It was assumed that the uncertainty, $\Delta x_{\nu}$ 
     in the position of the neutrino is related to its spread
    in momentum, $\Delta p_{\nu}$, by the relation:
     $\Delta x_{\nu}\Delta p_{\nu} \ge 1$. The requirement: $\lambda \Delta x_{\nu}
     = \ell_{osc}$, where $\ell_{osc}$ is the neutrino oscillation length, then
     leads to Eqn(6.1), it being assumed that if $\Delta x_{\nu} \simeq \ell_{osc}$
     it will be impossible to observe neutrino oscillations. Although this latter
      condition is evidently correct if $\Delta x_{\nu}$ refers to the experimental
      uncertainty in the observed position the neutrino detection event, it is easily
      shown to be completely
      false if, instead, it refers to the theoretical uncertainty calculated
      using the momentum-space Uncertainty Relation. For example, in the case of pion
      decay at rest
      the width of the coherent momentum wave packet associated with the spread in
      physical mass of the decay muon is $\Delta p = m_{\mu} \Gamma_{\mu}/m_{\pi}
      = 2.3 \times 10^{-16}$ MeV. Taking $\lambda = 10$, $\Delta m^2 = (1\rm{eV})^2$
      and $p_{\nu} = 29.8$ MeV,   
      Eqn(6.1) gives the limit $\Delta p_{\nu}/ p_{\nu}
       > 1.1 \times 10^{-14}$ as compared to the width of the coherent momentum 
       wave packet in pion decay at rest of $\Delta p_{\nu}/ p_{\nu}
      =  7.2 \times 10^{-18}$. This is three orders of magnitude lower than 
      Kayser's lower limit for the possibility of neutrino oscillations, (6.1).
       Yet explicit calculation of the corresponding damping of the interference
        term~\cite{JHF1}
       shows it to be completely negligible. As discussed in Reference~\cite{JHF1}
       the momentum spread of the neutrinos due to smearing of the physical mass of
       the decaying pion, or its thermal motion, are much larger than that associated
       with the physical mass of the decay muon. The two former sources are, however,
       incoherent, and so have no associated wave packets. 
      
      \par In fact, Kayser assumes that the mass eigenstates are both produced 
       and detected at the same times i.e. the equal velocity hypothesis is made,
       in contradiction with the different velocities resulting from the equal 
       momentum and unequal energy hypothesis. This procedure is justified by the
       hope that the effect of this logical inconsistency will be annulled by the
       space-time fuzziness introduced by the hypothetical wave packet:
       \par ` The wave packet treatment eliminates the need to make some
        idealising assumption by taking both momentum and energy variations
        properly into account'

        \par The `idealising assumption' referred to is actually the exact conservation 
        of energy and momentum that is automatic in all covariant calculations
        of decay or scattering processes. The real purpose of the 
        wave packet is, however, rather to enable the neutrinos, incorrectly
        assumed to be always produced at the 
        same time, but separating spatially due to their different velocities,
        to be detected at the same time. This is clearly impossible
        when, as is in fact the case, they move along essentially classical space-time 
        trajectories.
        
         \par Like Nussinov, Kayser introduces a hypothetical spatial wave packet of 
          length $c \tau$ where $\tau$ is the lifetime of the decaying state. This
          wave packet, which, as explained above, is only a classical
          wave theory analogue of the finite source lifetime, does not exist
          in the QM calculation. The Fourier transform of the exponential decay
          amplitude yields a Breit-Wigner amplitude describing the distribution of the
         physical mass of the unstable source particle. As discussed in the previous
         Section, as this mass is a property of the source particle, any resulting
         momentum
         smearing of the produced neutrinos is then an incoherent effect. The only
         coherent momentum smearing (yielding a momentum, not a spatial, wave packet)
         is that associated with the physical masses of any unobserved, unstable,
         recoil particles. Kayser then argues that the length of the spatial
          wave packet must be much shorter than  $c \tau$:
         \par `If we are interested in a neutrino emitted at time $t=0$, but
          we can learn only that the emitter was somewhere in a region of length
          $h$, then the amplitude for the emission to have ocurred at $t=0$ at the
          various points in this region must be added coherently. Thus the neutrino
          wave packet will have a length $d \simeq h$.'
         \par The false assumption is made that both neutrinos must be created at the 
          same time, $t=0$. It is then assumed, in contradiction with Eqn(5.1) above, 
          for the case of a neutrino source at rest, that the amplitudes of neutrinos
          created at different spatial positions must be added coherently. It corresponds 
          to performing the sums over $m$ and $l$ in Eqn(5.1) in a coherent manner, i.e.
          with $\sum_{m}\sum_{l}$ inside the modulus squared. An extended criticism of
          this assumption may be found in Reference~\cite{JHF1}. 
          In Kayser's interpretation
          the spatial wave packet is needed to `delocalise' the neutrinos, that, because
          of the assumption of a common production time, become spatially separated due to
           their 
          different velocities. In fact, the neutrinos can arrive at the detector at the
          same time because they may be produced at different times in the alternative
          histories corresponding to the different path amplitudes. There is then no
          need for the hypothetical spatial wave packet introduced by Kayser. There are
          no physical parameters in the QM calculation governing the size of such a wave packet.
         It does not exist except in an analagous classical wave theory calculation. The 
         space-time structure of particle propagation cannot be described by such a calculation. 
          \par Because equal velocities are assumed, i.e. that both neutrinos are 
           produced at one space time point and both detected at another, Kayser 
            obtains the standard oscillation phase, a result shown to be 
            unmodified~\cite{Kayser} by convolution with a narrow momentum wave packet.
           \par Ten years after Kayser's paper on wave packets in neutrino
            oscillations the first of many papers where spatial wave packets, as
            suggested by Kayser, were implemented using Gaussian functions, was
            written by Giunti Kim and Lee~\cite{GKL}. This paper starts with the
            statement:
            \par `If neutrinos are massive particles and mixed, a flavour 
             neutrino is created by a weak-interaction process as a coherent 
             superposition of mass eigenstates.'
             \par This statement is in contradiction with the findings of
             Shrock~\cite{Shrock1,Shrock2} that, if non-degenerate massive neutrinos
             exist, lepton flavour is not conserved by the weak interaction, so 
             that the different mass eigenstates are produced incoherently in
             different physical processes. It implies that, as previously assumed
             by Kayser (and all subsequent authors of papers on the QM of
            neutrino oscillations) that all neutrinos are produced at the same time.
            Therefore they must satisfy the equal velocity hypothesis of Section 2
            above, and independently of the introduction, or not, of spatial
            wave packets, or the assumption, or not, of exact energy-momentum
            conservation in the production process, the standard result for the
             oscillation phase must be 
            obtained.
            \par The introduction of Reference~\cite{GKL} contained a list of issues
             that it was claimed should be addressed in order to provide a `complete'
            treatment of the QM of neutrino oscillations. It is instructive to review
          this list from the perspective of the work presented in the present paper 
          and References~\cite{JHF1,JHF2}. The four issues were:
           \begin{itemize}
          \item[(i)]`A necessary condition for neutrino oscillations to occur is that
           the neutrino source and detector are localised within a region much
           smaller than the oscillation length; then the neutrino momentum has at
           least the corresponding spread given by the uncertainty principle
            ~\cite{Kayser}.'
         \item[(ii)]` The energy-momentum conservation in the process in which the
            neutrino is created implies that different mass-eigenstate components
           have different momenta as well as different energies~\cite{Win}.'
         \item[(iii)] `The different mass eigenstates must be produced and detected
          coherently; this is possible only if the other particles associated with
          the production and decay processes have energy momentum spreads larger
          than the energy momentum differences of the mass eigenstates.'
          \item[(iv)]`The wave function of the propagating neutrino must be a 
           superposition of the wavefunctions of the mass eigenstates with proper
           coefficients given by the amplitudes of the processes in which the mass
          eigenstate neutrinos are produced.'  
         \end{itemize}
           The following comments (C) may be made made on these points:  
         \begin{itemize}
         \item[(Ci)] The first sentence is trivially correct, but is unrelated 
          to the quantum mechanical aspects of the problem. The last phrase,   
          based as it is on the arguments of Kayser (Eqn(6.1)) is, as argued
          above, demonstrably without physical foundation.
         \item[(Cii)] This is indeed an essential ingredient of a correct treatment, 
          in QM, of neutrino oscillations. A corollary is that the neutrinos, 
          propagating over macroscopic distances, do so as classical
          particles~\cite{Moh1}. This condition is violated by the universal
          equal velocity assumption of Eqn(2.12).
         \item[(Ciii)]  Properly interpreted, the first sentence is correct.
          It means that the amplitudes describing the temporal sequence of
           events: a) propagation of the source, b) decay of the source,
           c) propagation of a neutrino mass eigenstate and d) production
            of the detection event must be added coherently. It does not mean
            that the mass eigenstates are part of a coherent `flavour eigenstate'
          in either the detection or production processes. The last phrase, as
           shown explictly for the case of pion decay at rest in Section 5
           above, is false.
        \item[(Civ)] The `wavefunction' referred to here is presumably a lepton
          flavour eigenstate propagating in space-time. As previously pointed
          out by Shrock~\cite{Shrock1,Shrock2} and shown later by Giunti, Kim
          and Lee themselves~\cite{GKL1}, such a state does not exist.
          \end{itemize}
           \par The effect of the Gaussian wave packet introduced in 
            Reference~\cite{GKL} on the oscillation phase is described in 
           Section 3 above. For a critical discussion of the `spatial coherence 
           length' and `momentum damping factor' generated by such wave packets
          see Reference~\cite{JHF1}. The point of view, explained there, of
          the present author is that, as the coherent spatial wave packet from
          which they are derived does not exist, they also are without any
          physical foundation. The oscillations are damped by the source 
         lifetime, as shown in Eqn(5.20), and also by the width of the
         coherent momentum wave packet associated with the different possible
         physical masses of unobserved recoil particles produced in the
         production process. As shown in Reference~\cite{JHF1}, both 
         effects are expected to be vanishingly small in any forseeable 
         neutrino oscillation experiment.
 \par An influential paper on the QM of heavy quark flavour oscillations was written by
  Lipkin~\cite{Lipkin1}. Similar ideas were applied to neutrino oscillations
  in Reference~\cite{GroLip} and in an unpublished pre-print~\cite{Lipkin2}.
  The starting point of Reference~\cite{Lipkin1} was the correct observation that
  all experiments measuring flavour oscillations actually observe only a spatially
  varying interference effect. This implies that all decay and detection times
  should be integrated over in order to derive the quantum mechanical probability
  to be compared with experiment. This is done, for example, in the derivation of
  Eqn(5.20) above. Lipkin interpreted this correct statement about the nature 
  of the experiments as implying that time should not appear at all in the
  theoretical description of the experiments. Formulae containing the time
   were referred to as describing `non experiments'. In all three papers
   cited above the initial state is incorrectly assumed to be in a pure
   flavour eigenstate that is a superposition of mass eigenstates. Thus,
   for example the $\rm{K}_S$ and the $\rm{K}_L$, or the different
   neutrino mass eigenstates are assumed to be produced at the same time.
   Since they must be detected at the same time, as only one detection
   event is observed, the equal velocity hypothesis (2.12) is thus 
   assumed. Since the additional assumption of equal energies and different
   momenta of the particles is made (of course in logical contradiction with
  the equal velocity hypothesis) the temporal part of the Lorentz
  invariant plane wave does not contribute to the oscillation phase.
  The latter is then entirely determined by the spatial part, as described in
  Section 4 above. The standard result for the oscillation phase is then
   obtained. The arguments given by Lipkin to justify the choice of equal
   energies and different momenta of the propagating mass eigenstates are
  unconvincing. In the process $\rm{K^-p \rightarrow \overline{K}^0 n}$
  Lipkin states that `Energy conservation requires the $\rm{\overline{K}^0}$
 to have a definite energy. When it is split into $\rm{K_L}$ and $\rm{K_S}$ 
   components with different masses the two states have the same energy
   but different momenta'. There would seem to be no physical justification
   for this apodictic statement. Momentum conservation requires the
  $\rm{\overline{K}^0}$ to be produced with a definite momentum. Why should
  it not then `split' into two states with the same momentum, and different
  energies? Later Lipkin gives an argument based on non-relativistic 
  kinematics (Eqn(2b) of Reference~\cite{Lipkin1}) as applied to the 
  process  $\rm{K^-p \rightarrow \overline{K}^0 n}$ to
  justify the neglect of the energy difference between the $\rm{K_L}$ and $\rm{K_S}$.
  Repeating the calculation using relativistic kinematics, as more appropriate
  to typical experimental conditions, shows instead that $\Delta p/ p = \Delta E/E$,
  so that there is, in this case, no kinematical justification for the equal energy
  hypothesis. Actually the $\rm{K_L}$ and $\rm{K_S}$, like the different neutrino
  mass eigenstates, are produced incoherently,
   in different physical processes, so that there is no production of the state
   `${\rm \overline{K}^0}$'. Conservation of energy and momentum then shows
   that they must have different energies, different momenta and different velocities. 
  \par In Reference~\cite{Lipkin1} a calculation of the oscillation phase for the
    $\rm{B^0 \overline{B}^0}$ system is performed in the laboratory
    system using temporal evolution, and allowing different propagation times
   for the different mass eigenstates (Eqn 9b of Reference~\cite{Lipkin1}).
   As discussed in Section 4 above, this corresponds to the full Lorentz invariant
   phase in the non-relativistic limit where $p_i \ll m_i$. In this limit
   the complete O($m^2$) Gribov-Pontecorvo result, with the oscillation
   phase a factor two larger than the standard result, is obtained. Lipkin
   noticed this difference but rejected the correct result 
   given, in the appropriate kinematical limit, by his Eqn 9b on
   the grounds that, as the time appeared explicitly in its
   derivation, it corresponded to a `non-experiment'.
   \par In Reference~\cite{GroLip}, similar arguments are applied to the neutrino
   oscillation case. Again, equal energies and different momenta and the
    (contradictory) equal velocity hypotheses are assumed, leading to the
    standard oscillation phase. The initial state is required to be
    a superposition of different mass eigenstates with pure flavour. 
     For example, in the case of say $\pi^+ \rightarrow \mu^+ \nu$, 
    the probability of detecting a $\nu_e$ is zero, and of detecting a 
    $\nu_{\mu}$ is unity, at $L = 0$. These conditions are used to
    fix the coefficients of the mass eigenstate superposition 
    at $L = 0$. However, as is shown by inspection of Eqn(5.21) above,
    exactly the same boundary condition is respected by the result
   of the Feynman path amplitude calculation where the $\pi^+$
    decays incoherently into the different mass eigenstates at 
    different times and no unphysical `flavour eigenstate wavefunction'
    is introduced. It was claimed in Reference~\cite{GroLip} that the 
    energy-momentum and space-time descriptions of flavour oscillations
    are `complementary' and that including them both leads to 
    `double counting' of the oscillation phase by a factor of two.
    In fact, the Lorentz invariant oscillation phase contains (except
    in the rest frame of the propagating particle) both
    energy-momentum and space-time contributions that must both 
    be included to obtain the correct result. The detailed considerations 
    of Sections 2-4 above show that attempts to use only energy-momentum
    or space-time descriptions in inappropriate kinematical regions
    leads instead to `half counting' of the correct Lorentz invariant
     phase.      
   \par In Reference~\cite{Lipkin2}, Lipkin justified the `equal energy' 
    hypothesis by reference to a paper by Stodolsky~\cite{Stodolsky} which 
    attempted a non relativistic density matrix description of flavour
    oscillations. It was proposed to use `stationary' beams of fixed energy
    to describe the system of propagating mass eigenstates. In this way the
    introduction of spatial wave packets was avoided. The present writer's
    opinion is that such an approach is ill-founded. In fact, the propagating
    particles describe classical trajectories, the detailed spatio-temporal
    structure of which is essential for the correct QM description of the
    phenomenon. This information is not available in the non-relativistic
    density matrix approach, which in any case,  is not appropriate to
    describe ultra-relativistic neutrinos. 
   \par The `stationary' source and target description with equal energies
    and different momenta for the neutrinos as well as Gaussian spatial 
    wave packets for both the source and detector was used more recently by Ioannisian
     and Pilaftsis~\cite{IP}. As the equal velocity assumption (2.12) was also
     made the standard oscillation phase was obtained.

    \par It is interesting to note that, in an earlier paper, written
     together with Kayser~\cite{KaySto} Stodolsky proposed a covariant 
     Feynman path amplitude approach, akin to that of the present paper
     and References~\cite{JHF1,JHF2}, to the description of `entangled'
     systems such as $\phi \rightarrow \rm{K_S} \rm{K_L}$. In this case,
     in contradiction to Reference~\cite{Stodolsky}, and as previously
     assumed by Kayser~\cite{Kayser}, equal momenta and different energies
     were proposed. The treatment of Reference~\cite{KaySto} differs from that of the
     present paper and References~\cite{JHF1,JHF2} in that the equal 
     velocity hypothesis (2.12) was assumed, so that the standard oscillation
     formula was obtained and the contribution to the oscillation phase of the
     coherent source was neglected. Also, unphysical spatial wave packets were
    introduced and the potentially important velocity smearing effects due to 
    variations of order $\Delta m_K$ in the physical mass of the $\rm{K_S}$,
    as discussed in Section 5 above, were neglected. The present writer is
    in agreement with the main conclusion of this paper, that the 
    `collapse of the wavefunction' often discussed in connection with entangled
    states, is only the collapse of a mathematical abstraction, that is actually
    irrelevant to the QM description of the experiment. It is perhaps, however,
    going a little too far to state, as in the last sentence of
     Reference~\cite{KaySto}
    that: `The best answer, finally to the `question of the collapse of the
    wavefunction' is that there is no wavefunction.' Indeed, the QM of 
    flavour oscillations is better described in terms of Feynman path
    amplitudes, as proposed in Reference~\cite{KaySto}, However, the wavefunction
     does remain an important 
    and useful concept in the description of `static' bound systems such as the
    hydrogen atom.
    \par Following Rich~\cite{Rich} several authors~\cite{KWeiss,GMS,GMS1} have
     used a non-relativistic Wigner-Weisskopf type formalism to describe the complete
     production-propagation-detection process. The propagating virtual neutrinos are 
     assigned
     the same energy and different momenta. In all cases the equal velocity
     assumption is made leading to the standard oscillation phase. The present
     writer's opinion is that such treatments take properly into account neither the
     ultra-relativistic nature of the propagating neutrinos nor the sequence
     of spatio-temporal production and detection events necessary for
     a correct calculation of the oscillation phase. In one paper using this
     non-relativistic approach~\cite{GMS1}, the effect of the source lifetime 
     was discussed. Although it was correctly concluded that the damping
     effect due to the finite source lifetime is negligible, it was proposed
     that the characterstic momentum spread in the Gaussian wave packet, 
     related to the source, for muon decay at rest (actually the Fourier transform 
     of a Gaussian spatial wave packet as proposed by Giunti Kim and Lee~\cite{GKL}
     following the suggestion of Kayser~\cite{Kayser}) should be  
     $\simeq 10^{-3}$ MeV. This is the typical momentum due to thermal motion
     at room temperature. As discussed above, any momentum smearing due to this 
     source or to the physical mass of the decaying muon is incoherent and not
     associated with any wave packet. Indeed, in the case of muon decay, since
     all recoil particles are stable, then, unlike in the case of pion decay,
     there is is no momentum wavepacket associated with neutrino propagation.
     \par The equal energy, different momentum, hypothesis was also made by the
      authors of Reference~\cite{KMOS}. This was justified by assuming that both
      production and detection of neutrinos resulted from inelastic scattering on
      an infinitely heavy target. Although such an assumption guarantees the
      kinematical correctness of the equal energy hypothesis, it evidently
      does not correspond to actual neutrino experiments where the neutrinos are
      produced by the decay of an unstable source, and where energy-momentum 
      conservation always requires (see Eqns(2.6),(2.7) above) that both 
      momenta and energies and, hence the velocities, are different. The equal
      velocity hypothesis (in contradiction with the different momenta of
      the neutrinos) was also made so that the neutrino propagator is purely
      spatial, giving, as shown in Section 4 above, the standard oscillation
      phase.
      \par Many of the features of the covariant path amplitude calculation of
      Section 5 above and References~\cite{JHF1,JHF2} have been previously 
      introduced into the discussion of neutrino oscillations. For example, the
      Lorentz invariant Feynman propagator for the neutrinos has been used in 
      References~\cite{Moh1,IP,Campagne,SWS,Shtanov,Beuthe}. The only author
     to introduce explicitly the invariant propagator of the source particle
      was Campagne~\cite{Campagne}. As the equal velocity hypothesis 
     was also made there is no contribution to the
     oscillation phase from this propagator and the standard result was 
     obtained for the oscillation phase.
     \par The authors of Reference~\cite{LDR} recognised that the derivation
      of the standard formula required the equal velocity hypothesis, and that
       if exact energy-momentum conservation is imposed, so that the neutrinos
       have different
      times-of-flight, an oscillation phase a factor of two larger is obtained.
      In spite of noting that neutrinos of widely differing masses, as
      expected theoretically, cannot have equal velocities, the use of the
      equal velocity hypothesis was, nevertheless, recommended. A short
      note of Okun and Tsukerman~\cite{OT} pointed out the kinematical
      impossibility of the equal velocity assumption for neutrinos
      with different masses.
       \par A paper by Giunti~\cite{Giunti} considering the analogy between
       the interference effects in the Young double slit experiment and in
       neutrino oscillations, claimed to demonstrate that the extra factor
       of two in the oscillation phase, due to neutrino propagation, obtained
       when the different neutrino velocities are correctly taken into
       account~\cite{JHF1,LDR}, is incorrect. This argument was based
       on the correct observation that in the Young double slit experiment
       with photons, photon propagation gives no contribution to the 
       interference phase. That this must be so is evident by setting
        $m_i = 0$ in Eqn(5.9) above. Giunti then claimed to have rejected
        the different velocity hypothesis~\cite{LDR} and the path amplitude
         calculations
        of Reference~\cite{JHF1} by {\it reductio ad absurdum} since, in the
        analagous Young double slit experiment, a vanishing interference
        phase, clearly excluded by experiment, is apparently predicted.
         However, Giunti neglects the
        contribution to the interference phase of the coherent 
        source (the excited atom that produces the photon). The contribution of
        the source
        to the interference phase is, using the formula analagous to Eqn(5.7)
        above, for an atomic radiative transition:
        \begin{equation}
        \Delta \phi^{source} =  E^{\star}(t_1-t_2) \simeq E_{\gamma}(t_1-t_2) =
         E_{\gamma}(r_2-r_1)
        \end{equation}
         where $E^{\star}$, and $E_{\gamma}$ are respectively the atomic
         excitation energy and the photon energy, while $t_1,t_2$ and 
         $r_1,r_2$ are defined in Reference~\cite{Giunti}. This formula
         is identical to Eqn(8) of Reference~\cite{Giunti}, and since
         the contributions from the photon propagators vanish, the path
         amplitude calculation gives the usual result obtained in the
         classical wave theory of light. Having wrongly concluded that the
         photon is produced at a unique time, Giunti introduces a 
         hypothetical Gaussian spatial wave packet (unphysical, because
         it has no relation to the physics of the photon production
         process) and demonstrates that the space-time smearing provided
         by the wave packet allows to recover the same interference phase
         as in Eqn(6.2) above. Also the interference term is found to be
         damped by a factor dependent on the length of the wave packet.
         As discussed in Reference~\cite{JHF1}, as well as above in the present 
         paper, this damping factor, derived from the spurious spatial
         wave packet is also without any physical foundation. The
         damping of the interference term is actually produced by the
         finite lifetime of the coherent source that limits the value
         of $t_1-t_2$ in Eqn(6.2).
      \par A recent paper by De Leo, Nishi and Rotelli~\cite{DLNR} has
       also considered the effect on the oscillation phase of different
       kinematical assumptions. As in Reference~\cite{LDR} it was realised
       that only in the case of the equal velocity hypothesis is the 
       standard oscillation phase obtained; in any other case the phase 
       is a factor of two larger. This agrees with the conclusions of
       Sections 2, 3 and 4 of the present paper. The difference between
       the kinematical discussions of the present paper and those of
       Reference~\cite{DLNR} is that in the former case, in accordance
       with the experimental conditions, and as previously pointed out
       in Reference~\cite{Lipkin1}, a constant distance, $L$, is assumed
       between the source and the detection event. In the latter case
       this distance is allowed to vary, and also different creation
       times are allowed for the mass eigenstates. In the general kinematical
       discussion of Section II of Reference~\cite{DLNR}, which may be 
       compared to Section 2 above (since both discuss the Lorentz invariant
       oscillation phase) it is assumed that $\Delta t$ is the same,
       but that $\Delta x$ is different 
       for the different neutrinos. This is in complete disagreement with
       the experimental conditions of typical neutrino oscillation
       experiments. In order to permit different source-detector
       separations, a Gaussian spatial wave packet was
       introduced. It was concluded that by a suitable choice of creation
       times a pure flavour eigenstate can be obtained at creation. Indeed
       the necessary existence of such a state is the initial hypothesis on
       which the arguments given in the paper are based. It may be commented
       that, firstly, as pointed out long ago by Shrock~\cite{Shrock1,Shrock2}
       no such flavour eigenstate exists since the neutrinos are created
       in separate, incoherent, processes, and secondly, the distance $L$ must be
       constant, since the source is assumed to be at rest. Finally, although it
       is true that the different
       neutrinos may be created at different times, the 
       spatial wave packet introduced to allow the possibility of 
       different source-detector distances does not, as argued above,
       have any physical basis.
       \par A very recent paper by Beuthe~\cite{Beuthe} makes the same basic 
       assumptions as an earlier paper by Giunti {\it et al}~\cite{GKLL}.
       Both equal energy and equal energy hypotheses are considered, but the
       equal velocity condition (2.12) is always assumed so that the standard 
       oscillation phase is always recovered, and the important contribution 
       to the oscillation phase of the coherent source particle is neglected. There
       is a lengthy discussion of the effects of hypothetical
       Gaussian wave packets used
       to describe both the source and detector particles.
       \par Further critical discussion of different treatments in the literature
        of the QM of neutrino oscillations may be found in Reference~\cite{JHF1}.
        The same paper also describes briefly two atomic physics experiments,
        the `quantum beat' experiment~\cite{BerSub} and the `photodetachment
        microscope'~\cite{BDD,BBGKM,BBD} where the Feynman path amplitude 
        description has been successfully tested in experiments where spatially
        varying interference effects, very similar to particle flavour 
        oscillations, have been observed.      

 \SECTION{\bf{Summary and Outlook}}
    
    The kinematical and geometrical discussion of Sections 2, 3 and 4 above
  shows that the standard formula (1.5), for the contribution of neutrino
  propagation to the oscillation phase, is a consequence of the equal
  velocity hypothesis where it is assumed that both interfering neutrinos are 
  produced at the same space-time point. This hypothesis is incompatible with
  the propagation of the neutrinos along classical space-time trajectories
  if they have different masses. The oscillation phase, calculated at O($m^2$),
   using the equal velocity hypothesis is found to be a factor of two smaller
   than the result first obtained by Gribov and Pontecorvo~\cite{GribPont}.
    The latter is obtained by imposing both exact energy-momentum conservation
    and a consistent geometrical propagation in space-time,
    taking properly into account the
    different neutrino velocities. It is referred to above as the `exact'
   O($m^2$) formula. Thus the standard formula neglects numerically important
    O($m^2$) terms as compared to the exact one.
    \par This conclusion remains the same whether the neutrino propagator is
    described in a Lorentz invariant manner using plane waves (Section 2),
    whether convolution with a Gaussian wave packet is performed (Section 3)
    or whether only temporal or spatial evolution of the neutrino wave function
     is considered (Section 4). In contrast, the assumptions of equal momenta
    and different energies or of equal energies but different momenta give
    only negligible  O($m^4$) corrections to the exact formula, or to the
    standard formula, as compared to the phase calculated assuming
    exact energy-momentum conservation.

    \par The Feynman path amplitude calculation of Section 5 is based on 
     the use of the exact Lorentz invariant plane wave formula
     (or, equivalently, the invariant Feynman space-time propagator) for the
      contribution
     of neutrino propagation to the oscillation phase. Since the different
     neutrino mass eigenstates are produced in independent physical 
     processes~\cite{Shrock1,Shrock2,JHF1} the decay of the source can 
     occur at different times in the amplitudes describing the 
     propagation of different neutrinos. The oscillation phenomenon
     then occurs respecting the constraints of both exact energy-momentum
     conservation and exact space-time geometry. The different decay times
    of the source in the interfering amplitudes then lead to an 
    important contribution to the oscillation phase from the space-time
    propagator of the source.
    \par Damping of the interference term due to the finite source lifetime
    or momentum smearing related to the off-shell nature of the source or
    recoil particles has been previously found to be completely negligible in all 
    foreseeable neutrino oscillation experiments~\cite{JHF1}.
    \par All previous calculations in the literature in which the standard
     oscillation phase is obtained assume the production of 
     a coherent `neutrino flavour eigenstate' in the decay process, that is
     a quantum superposition of mass eigenstates. This enforces equal 
     production times for the all mass eigenstates and hence the equal 
     velocity condition. As pointed out in References~\cite{Shrock1,Shrock2,GKL1,JHF1}
     such a `neutrino flavour eigenstate' does not exist. The different
     neutrino mass eigenstates are produced incoherently in different physical
     processes. This universal, incorrect, assumption concerning the nature of the
      initial state of the neutrinos thus explains why the standard 
     oscillation phase has been, hitherto, universally obtained. Thus the 
     contribution to the oscillation phase of neutrino propagation has been
     generally underestimated by a factor of two, and the important contribution
     to the phase of the coherent source (resulting from different source 
     decay times in the interfering amplitudes) has been universally neglected.
     \par As discussed in some detail in Section 6, in order to enable the 
      interference phenomenon, leading to `neutrino oscillations' to occur
      when the different mass eigenstates are produced at the same time 
      inside the `neutrino flavour eigenstate' it was proposed~\cite{Kayser}
      to introduce a spatial wave packet to delocalise the neutrinos. Moving
      with different velocities along classical trajectories, and produced
      at the same space time-point, it is clear the neutrinos can never 
      arrive together at the unique space-time point of the detection
      event. The spatial `fuzziness' introduced by the hypothetical
      wave packets, associated with each neutrino, was conjectured to enable them both
      to have non-vanishing amplitudes at the position of the 
      detection event, so that neutrino oscillations can occur. In fact,
      because the interfering neutrinos can be produced at different times
      there is no need, to produce neutrino oscillations, for the spatial fuzziness
      introduced by the hypothetical wave packets.

      \par The hypothetical wave packet suggested in Reference~\cite{Kayser}
       does not exist in the QM calculation of the neutrino production
       process. The neutrinos (particles) are produced in space-time 
       according to an exponential decay law. A `wave packet' is only a
       (very loose) analogy in classical wave theory to the effect of the
       lifetime of an unstable particle on the energy (or mass)
       distribution of its decay products. In the analogy this distribution
       is given by the Fourier transform of the spatial wave packet, but
       no there is no such wave packet in the QM calculation itself. Thus the
       (mathematically convenient) Gaussian wave packets that abound in the
       literature on the QM of neutrino oscillations have no physical
       foundation within QM. The relevant (related) parameter, in
       the quantum mechanical calculation, is the mean lifetime of the source. 
       There is no physical connection between this parameter and the
       length of a loosely analagous wave packet with an arbitary Gaussian
       form. The introduction of such wave packets in the QM calculation
       mixes up in a confused way concepts from QM and a classical wave
       theory that contains no information on the space-time evolution
       of particle positions.
 \begin{figure}[htbp]
\begin{center}
\hspace*{-0.5cm}\mbox{
\epsfysize15.0cm\epsffile{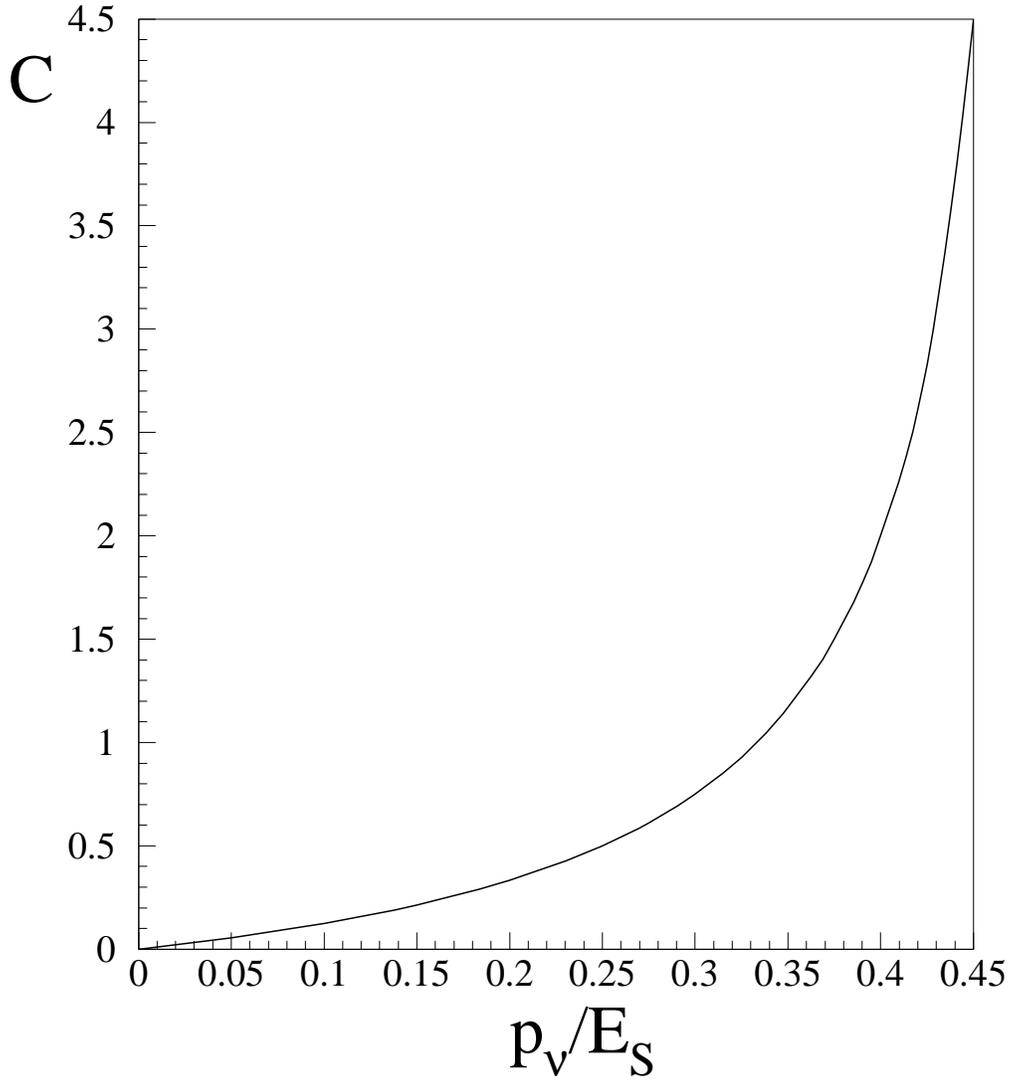}}
\caption{Correction factor for $\Delta m^2$ relating the standard formula
  to the Feynman path amplitude calculation.}
\label{fig-fig1}
\end{center}
\end{figure}
       \par The most important additional features of the present paper as compared
       to References~\cite{JHF1,JHF2} are, first, the realisation that the standard
       oscillation phase follows {\it only} from the manifestly unphysical equal
       velocity hypothsis and is quite unrelated to the use of Gaussian wave packets
       or any of the other kinematical assumptions made in the derivations. The misleading
       impression may have been given in References~\cite{JHF1,JHF2} that the
       standard oscillation phase, when obtained in covariant calculations,  
       was a consequence of the use of wave packets. The second is the
       realisation that the incoherent nature of the neutrino production process,
       which is the physical basis of the calculations presented in References
       ~\cite{JHF1,JHF2} and Section 5 above, was aleady pointed out more than twenty years
       ago by Shrock~\cite{Shrock1,Shrock2} in the published literature. I was not aware of this
       work at the time of writing References~\cite{JHF1,JHF2}. 
       \par In closing a few remarks are made on phenomenology and experimental tests.
         The mass difference, $\Delta m_{stand}^2$, derived from experimental results using
        the standard formula, is related to that, $\Delta m_{FP}^2$, given by the 
        Feynman path amplitude calculation, by the formula (valid for any source at rest):
        \begin{equation}
         \Delta m_{FP}^2 \equiv C \Delta m_{stand}^2 =
          \frac{\Delta m_{stand}^2}{\frac{E_S}{p_{\nu}}-2}
        \end{equation}
      while for neutrino oscillations following two body decays in flight of ultra-relativistic
      charged pions or kaons~\cite{JHF1}:
     \begin{equation}
         \Delta m_{FP}^2 = \frac{(1-R_m^2)\Delta m_{stand}^2}{2 R_m^2}
        \end{equation}
       where $R_m$ is defined after Eqn(2.8). In the case of experiments involving neutrino
      production in pion, kaon or muon decay the conversion is straightforward. However,
      for neutrinos produced in nuclear reactors the effective oscillation phase will
      require a suitable average, with appropriate weighting factors, over the decays of 
      all $\beta$-unstable nuclei contributing to the neutrino flux. For a given nuclear
      species the correction factor, $C$, to the standard oscillation phase is given,
       by Eqn(7.1),
      as $1/(E_{\beta}/p_{\nu}-2)$, where $E_{\beta}$ is the total energy release in the 
      decay. This is evidently an immense undertaking for any actual reactor-based experiment. 
      Any phenomeological conclusions hitherto drawn from the results of such experiments, 
      using the standard formula, must therefore be discarded if the 
      oscillation phase is correctly given by the path amplitude calculation.
      \par Of particular interest, in view of the recent results of the Kamiokande
       ~\cite{Kamiok} and SNO~\cite{SNO} collaborations, are the $\beta$-decay
      processes: $\rm{^8B} \rightarrow \rm{^8Be}^{\ast}+e^++(\nu_1,\nu_2,\nu_2)$
      contributing to the flux of high energy solar neutrinos. The correction
      factor, $C$, in Eqn(7.1), is plotted in the range: $0<p_{\nu}/E_S<0.45$ in
      Fig.1. For $\rm{^8B}$ $\beta$-decay, $E_S \simeq 28$ MeV. The correction
       factor is unity when $p_{\nu}/E_S = 1/3$, corresponding to 
       $p_{\nu} \simeq 9.3$ MeV. Near the kinematical end-point at
       $p_{\nu} \simeq 14$ MeV, $C$ rises steeply, reaching a maximum value
       of about 1500. Thus neutrino oscillations, observable when $C \simeq 1$, are
       strongly suppressed\footnote{i.e. they correspond to a vanishing mass 
        difference or an infinite wavelength for the oscillation in the standard 
        formula.}
 in the near end-point region. For the case of the electron
       capture reactions:  $e^-+\rm{^7Be} \rightarrow \rm{^7Li}+(\nu_1,\nu_2,\nu_2)$
       contributing line spectra to the solar neutrino flux, $p_{\nu} \simeq E^{\ast}$
       where $ E^{\ast}$ is the excitation energy of the unstable $\rm{^7Be}$ atom and
       $C = -1$, so that the oscillation phase of the path amplitude calculation
       is the same as that given by the standard formula.
      \par As already mentioned in Reference~\cite{JHF1}, evidence for neutrino 
       oscillations in short baseline experiments such as LNSD~\cite{LNSD} and
       KARMEN~\cite{KARMEN} can be confirmed or invalidated by a search for muon 
       oscillations following pion decay at rest, since, the muon oscillation 
       phase is found~\cite{JHF1} to be identical to that of the associated
       neutrinos given in Eqn(5.21) above. Evidently the event rate in such muon
       oscillation  experiments can exceed that possible in the search for the
       associated neutrino  oscillations by many orders of magnitude.

\end{document}